\documentclass[aps,prx,twocolumn,showkeys,noprintnumbers,amssymb,aps,superscriptaddress,longbibliography]{revtex4-2}
\usepackage{graphicx}% Include figure files
\usepackage{bm}% bold math
\usepackage{color}
\usepackage{amsmath}
\usepackage[breaklinks,colorlinks,bookmarks=false,citecolor=blue,linkcolor=red,urlcolor=blue]{hyperref}

\newcommand{\be}{\begin{equation}}
\newcommand{\ee}{\end{equation}}
\newcommand{\bfs}{\bm{s}}
\newcommand{\Neel}{N\'eel}
\newcommand{\llangle}{\langle\!\langle}
\newcommand{\rrangle}{\rangle\!\rangle}
\begin{document}

%%%% Publication information
\onecolumngrid
\vspace*{-11.5mm}

\rightline{published in \href{https://doi.org/10.1515/zna-2026-0016}{Z.\ Naturforsch.\ A {\bfseries 81}, 531 (2026)}}
\vspace*{2mm}
\twocolumngrid
%\advance\textheight by 2 true mm
%%%%

\title{Field-induced states and thermodynamics of the frustrated
Heisenberg antiferromagnet on a square lattice}

\author{Andreas Honecker}

\affiliation{Laboratoire de Physique Th\'eorique et
Mod\'elisation, CNRS UMR 8089, CY Cergy Paris Universit\'e, Cergy-Pontoise, France}

\author{M. E. Zhitomirsky}

\affiliation{Universit\'e Grenoble Alpes, CEA, IRIG, PHELIQS, 38000 Grenoble, France}

\author{Alexander Wietek}

\affiliation{Max-Planck-Institut f\"{u}r Physik Komplexer Systeme,
        N\"{o}thnitzer Stra{\ss}e 38, 01187 Dresden, Germany}

\author{Johannes Richter}

\affiliation{Max-Planck-Institut f\"{u}r Physik Komplexer Systeme,
        N\"{o}thnitzer Stra{\ss}e 38, 01187 Dresden, Germany}

\affiliation{Institut f\"ur Theoretische Physik,
  Otto-von-Guericke-Universit\"at Magdeburg,
  39016 Magdeburg, Germany}

\date{January 12, 2026; revised February 10, 2026}

\begin{abstract}
We investigate the ground-state and finite-temperature properties of the $J_1$-$J_2$ Heisenberg antiferromagnet on the square lattice in the presence of an external magnetic field. We focus on the highly frustrated regime around $J_2 \approx J_1/2$. The $h$-$T$ phase diagram is investigated with particular emphasis on the finite-temperature transition into the ``up-up-up-down'' state that is stabilized by thermal and quantum fluctuations and manifests itself as a plateau at one half of the saturation magnetization in the quantum case.
We also discuss the enhanced magnetocaloric effect associated to the ground-state degeneracy that arises at the saturation field for $J_2=J_1/2$.
For reference, we first study the classical case by classical Monte Carlo simulations.
Then we turn to the extreme quantum limit of spin-1/2 where we perform zero-
and finite-temperature Lanczos calculations.
\end{abstract}

\maketitle

\section{Introduction}

The
Heisenberg antiferromagnet with nearest-neigh\-bor,
$J_1 > 0$, and next-nearest-neighbor
bonds, $J_2\geq 0$, called
$J_1$-$J_2$ model, or frustrated square lattice antiferromagnet (FSAFM),
is the archetypical frustrated spin model \cite{Lhuillier2001,Lacroix2011}.
The corresponding Hamiltonian is given by
\begin{equation}
\label{eq:ham}
 {\cal H} =  J_1\sum_{\langle i,j \rangle} \bfs_i \cdot \bfs_j + J_2\sum_{\llangle i,j
 \rrangle}\bfs_i \cdot \bfs_j - h\sum_i s^z_i
\end{equation}
including a coupling with an applied magnetic field $h$. Here,
$\langle \cdot, \cdot \rangle$ denotes summation over the nearest-neighbor bonds $J_1$
whereas $\llangle \cdot, \cdot \rrangle$ corresponds to the next-nearest neighbor $J_2$
bonds that couple spins on the diagonals of  square plaquettes.

The $J_1$--$J_2$ square-lattice model was introduced in the late 1980s
with the aim to describe the breakdown upon doping of the \Neel\ antiferromagnetic  (NAF)
state in the high-temperature cuprate superconductors
\cite{inui88,chandra88,dagotto89}.
Over the last few decades the zero-field case, $h=0$,  has attracted much
attention, since  it is a canonical model to
study frustration-induced quantum phase transitions between semiclassical
ground-state phases with a magnetic long-range order (LRO) and  disordered quantum
phases, see, e.g.,
Refs.~\cite{inui88,chandra88,dagotto89,schulz92,schulz96,richter93,richter94,
zhito96,Trumper97,bishop98,singh99,capriotti01,sushkov01,rgm_ihle,Sir:2006,Schm:2006,mambrini2006,darradi08,ortiz,cirac2009,ED40,fprg,Schmidt2011,yu2012,balents2012,becca2012,verstrate2013,becca2013,beach2013,gong2014,doretto2014,Ren2014,wang2014,chou2014,ccm2015,Wang2016,Reuther2017,Sandvik2017,Sheng2017,Poilblanc2017,Haghshenas2018,Liu2018,Zhao2020,Liu2020,Hasik2021,Imada2021,Liu2022,Liu2022a,Roth2023,Qian2024,Ruckriegel2024,Lin2024,Huang2024,Vecsei2025,Qian2025,Jin2025,qiao2025}.

Based on numerous studies using a plethora of quantum many-body methods,  consensus
has been achieved that the $J_1$-$J_2$ model in the extreme quantum limit
$s=1/2$ exhibits a magnetically disordered
spin-rotation-invariant
quantum
phase in a small region $J_2^{c1} < J_2 < J_2^{c2}$ around $J_2 = 0.5J_1$.
The semi-classical ground-state phases of the model, namely
the N\'{e}el antiferomagnetic  LRO at
small $J_2/J_1 < J_2^{c1}$
and the  collinear antiferromagnetic LRO  at large $J_2/J_1>J_2^{c2}$, are
well understood.
Concerning the critical values $J_2^{c1}$ and $J_2^{c2}$,
some estimates are $J_2^{c1}=0.44-0.46$ and
$J_2^{c2}=0.59-0.6$
\cite{darradi08,gong2014,chou2014,ccm2015,Sandvik2017,Liu2018,Hasik2021,qiao2025}.
However, some recent PEPS approaches \cite{Wang2016,Poilblanc2017,Sheng2017,Qian2024}
lead to a transition point $J_2^{c1}$  close  to $0.5$
or even above.
Concerning the nature of
the intermediate magnetically disordered quantum phase,
there is an ongoing active  controversial debate
\cite{Sir:2006,darradi08,yu2012,balents2012,becca2012,verstrate2013,becca2013,beach2013,gong2014,doretto2014,Ren2014,wang2014,chou2014,ccm2015,Wang2016,Reuther2017,Sandvik2017,Sheng2017,Poilblanc2017,Liu2018,Zhao2020,Liu2020,Imada2021,Liu2022a,Qian2024,Jin2025}.
A particular controversy concerns
the very existence of an
excitation gap in the intermediate quantum phase.
A finite excitation gap was reported
in Refs.~\cite{ED40,balents2012,becca2012,doretto2014}, whereas
indications of a gapless
spin liquid state were found in Refs.~\cite{verstrate2013,becca2013,chou2014,Liu2022a}.
Investigations based on the density matrix renormalization group
 with explicit implementation of $SU(2)$ spin
rotation symmetry
in Ref.~\cite{gong2014} report on a gapless spin liquid for $0.44 < J_2/J_1
< 0.5$ and a gapped plaquette valence bond phase for  $0.5 < J_2/J_1 <
0.61$.
This structure of the phase diagram is reaffirmed in Ref.~\cite{Liu2022a}, but the point separating the two
intermediate phases is shifted to $J_2/J_1 \approx 0.56$. On the other hand,
a recent matrix-product state method claims that there is no spin-liquid phase in this model \cite{Qian2024}.

In addition to the basic theoretical interest in this nontrivial  quantum many-body model,
it also attracts attention thanks to
its relation to experimental studies  of various magnetic materials,
such as VOMoO$_4$ \cite{VOMoO4},
$\mathrm{Li}_{2}\mathrm{VOSiO}_4$ and $\mathrm{Li}_{2}\mathrm{VOGeO}_4$ \cite{melzi00,melzi01,rosner03},
BaCdVO(PO$_4$)$_2$ \cite{Povarov2018},
{RbMoOPO}$_{4}${Cl}  \cite{Takeda2021},
{SrLaCuSbO}$_{6}$ and {SrLaCuNbO}$_{6}$ \cite{Watanabe2022},
{NaZnVOPO}$_{4}$({HPO}$_{4}$) \cite{Guchhait2022},
{YbBi}$_{2}${IO}$_{4}$ and {YbBi}$_{2}${ClO}$_{4}$ \cite{Park2024},
and
K$M$PO$_{4}$ \cite{Fujihala2025,Xu2025}.
However, to the best of our knowledge so far no magnetic  material with  an exchange-parameter ratio
$J_2/J_1$
suitable for a magnetically disordered ground-state phase has has yet been
found.

The unconventional quantum nature of the zero-field ground state in the
strong-frustration regime has also stimulated the search for unconventional
features in the magnetization curve
\cite{Yang1997,Honecker2000a,Honecker2000b,Honecker2002,Schulenburg2002,Chern2002,Jackeli04,Shannon2006,Thalmeier08,Coletta2013,Shibata2016,Yamaguchi2018,Sur2024}
as well as finite-temperature properties \cite{rgm_ihle,Shannon2004,Schmidt2005,Schmidt06,Schmidt07a,Schmidt07b,Seabra2009,Mikheyenkov2016,Schmidt2017,Prelovsek2020,Gauthe2022} of the spin-1/2 FSAFM.
In particular,
in the regime of strong frustration $J_2/J_1 \approx 0.5$, where the zero-field
quantum ground state does not exhibit magnetic LRO, evidence for a well-pronounced plateau
at $1/2$ of saturation was found \cite{Honecker2000a}.

In this paper, we revisit the behavior of the frustrated square-lattice antiferromagnet
in an external magnetic field at zero and at finite temperature, where we focus on the
regime around $J_2/J_1=0.5$.
In Sec.~\ref{sec:Class} we first revisit the $h$-$T$ phase diagram of the
classical Heisenberg model at $J_2/J_1=0.5$. Then, in Sec.~\ref{sec:Quantum} we switch to the
extreme quantum limit of spin $s=1/2$. We first revisit the zero-temperature phase diagram
in an external field in Subsec.~\ref{sec:qGS} and then study finite-temperature properties
in an external field in Subsec.~\ref{sec:qT}.
Finally, we conclude in Sec.~\ref{sec:Concl} with a summary and perspectives.
We note that while much of the work on the spin-1/2 model focusses on the controversial
$h=0$, $T=0$ phase diagram in the regime around $J_2/J_1=0.5$, here we do \emph{not} intend to contribute
to this particular question that is intensively studied by many other authors, as reviewed above.

\section{Classical model}
\label{sec:Class}

\subsection{Ground states}
\label{sec:ClassGS}

We begin with the classical $J_1$-$J_2$ model (\ref{eq:ham}) treating spins $\bfs_i$ as  
three-component unit  vectors. For the special ratio of the two exchange constants $J_2/J_1= 1/2$, 
magnetic frustration is enhanced featuring an infinite classical degeneracy for all magnetic fields
below the saturation \cite{Honecker2000a}. In order to see this we define the  total spin of a square plaquette $\alpha$:
\begin{equation}
\bm{L}_\alpha =  \sum_{i \in \alpha} \bfs_i \, ,
\label{eq:PlaquetteSpin}
\end{equation}
as a sum over four surrounding sites.
Then,  for $J_2=J_1/2$, the spin Hamiltonian (\ref{eq:ham}) can be rewritten as a function of
$\bm{L}_\alpha$ variables only
\begin{equation}
 {\cal H} =  \frac{1}{4}\sum_{\alpha} \Bigl( J_1 \bm{L}_\alpha^2 - 
 h L^z_\alpha \Bigr) .
\label{eq:hamPlaquette}
\end{equation}
So far, this is an exact transformation valid for any value of the spin. Let us now focus on
 the classical case. Minimizing energy of a single square plaquette we obtain from 
 Eq.~(\ref{eq:hamPlaquette}):
 \begin{equation}
L^z_\alpha = h/(2J_1) \,.
\label{Lz}
 \end{equation}
Any classical spin configuration that satisfies the  above constraint   for {\it every} square plaquette yields the lowest ground-state  energy. The subsequent analysis of degenerate ground states
splits into two parts: (i) finding equilibrium four-sublattice configurations for one block
and (ii) extending  single-block states to the entire lattice. Before doing that we note 
that the constraint (\ref{Lz}) gives the following  equilibrium magnetization 
 $m = L^z_\alpha/4 =h/(8J_1)$, which in turn yields for the saturation field
 $h_{\rm sat}=8J_1$. At $T=0$,  the classical magnetization
curve $m(h)$   is a  straight line up to  $h_{\rm sat}$.

\begin{figure}[t]
\begin{center}
\includegraphics[width=0.9\columnwidth,angle=0]{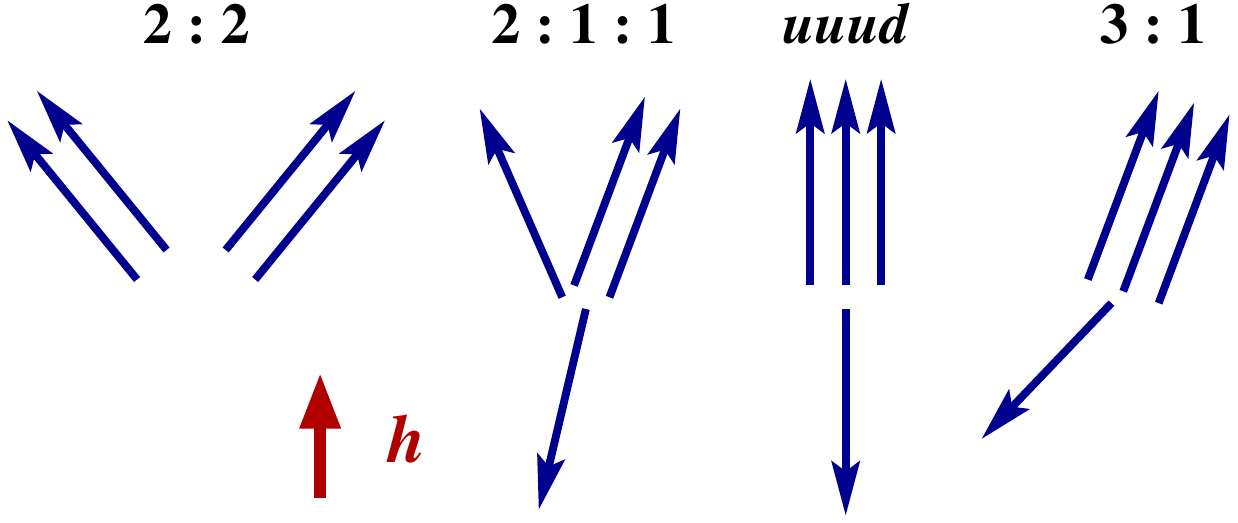}
\end{center}
\caption{Spin configurations minimizing the classical energy of a single four-site plaquette
of the FSAFM in different magnetic fields. The collinear $uuud$ state produces the 1/2 magnetization plateau.
}
\label{SConf}
\end{figure}

The constraint  (\ref{Lz}) imposes three conditions on the eight angle variables for the four spins
of each block. One more angle variable is eliminated by the broken $O(2)$ rotation symmetry about
the field direction. Overall, this counting yields four continuous variables to describe various 
classical ground states for a single block. Analysis of the effect of fluctuations, thermal or quantum,
on a state selection can be performed using the concept of an effective biquadratic exchange,
see, for example, Ref.~\cite{Zhitomirsky2014}.
Figure \ref{SConf} sketches the relevant spin configurations that appear for four-site blocks
in various magnetic fields
\cite{Honecker2000a,Heinila1993,Penc2004}. The ``up-up-up-down'' ($uuud$) configuration has the lowest energy
for magnetic fields around $h_{\rm sat}/2$ and produces the magnetization plateau with $m=1/2$. 
At higher fields away from the 1/2 plateau, spins tilt retaining the 3:1 splitting. For fields below $m<1/2$,
the noncollinear spin configuration has a more complex  2:1:1 structure. The 2:2 state, which is one of the classical ground states
in the entire range $0<h<h_{\rm sat}$, is stable, in the presence of a biquadratic exchange, in the low-field region only
\cite{Penc2004}.

Since the plaquettes $\alpha$ share edges, one spin participates in several plaquettes. As a result, one faces the problem
of extending a
single-block solution to the entire lattice. This leads to additional degeneracy  for the lattice spin model for each of
the single-block solutions in Fig.~\ref{SConf}. The corresponding degeneracy is most easily visualized for 
 the $uuud$ state  \cite{Honecker2000a}. Specifically,
consider the most symmetric  $uuud$  configuration. For that we choose the down spin, for example,
 in the lowest left corner of the first square plaquette and extend this configuration in a translationally invariant way over the 
 whole lattice. This state consists of parallel, vertical and horizontal, lines of
up-down spins separated by lines of up spins. The alternating up-down spins along any of these lines can be freely rotated around
an arbitrary angle without changing the total classical energy
\cite{Honecker2000a}. Since the $uuud$ state permits a maximum of such
deformations, one may expect it to be selected by thermal fluctuations via the ``order-by-disorder''
mechanism \cite{Villain1980,Shender1982,Rastelli1983,Kawamura1984,Henley1989,Shender1996}.
A subtle property of the order-by-disorder mechanism is a competition between a local fluctuation contribution
manifested by the effective biquadratic coupling and the extended zero-energy modes related to spin distribution
over the entire lattice. Furthermore, in the latter case there is the possibility of distinct  effects  from 
the thermal fluctuations, which rely on the entropic contribution of the  low-energy (zero)  modes, and from 
the  quantum fluctuations, with magnons at different momenta contributing equally.
As we shall see below, all these exotic features of the order-by-disorder effect are relevant for the FSAFM.

\begin{figure}[tb]
\begin{center}
\includegraphics[width=0.95\columnwidth,angle=0]{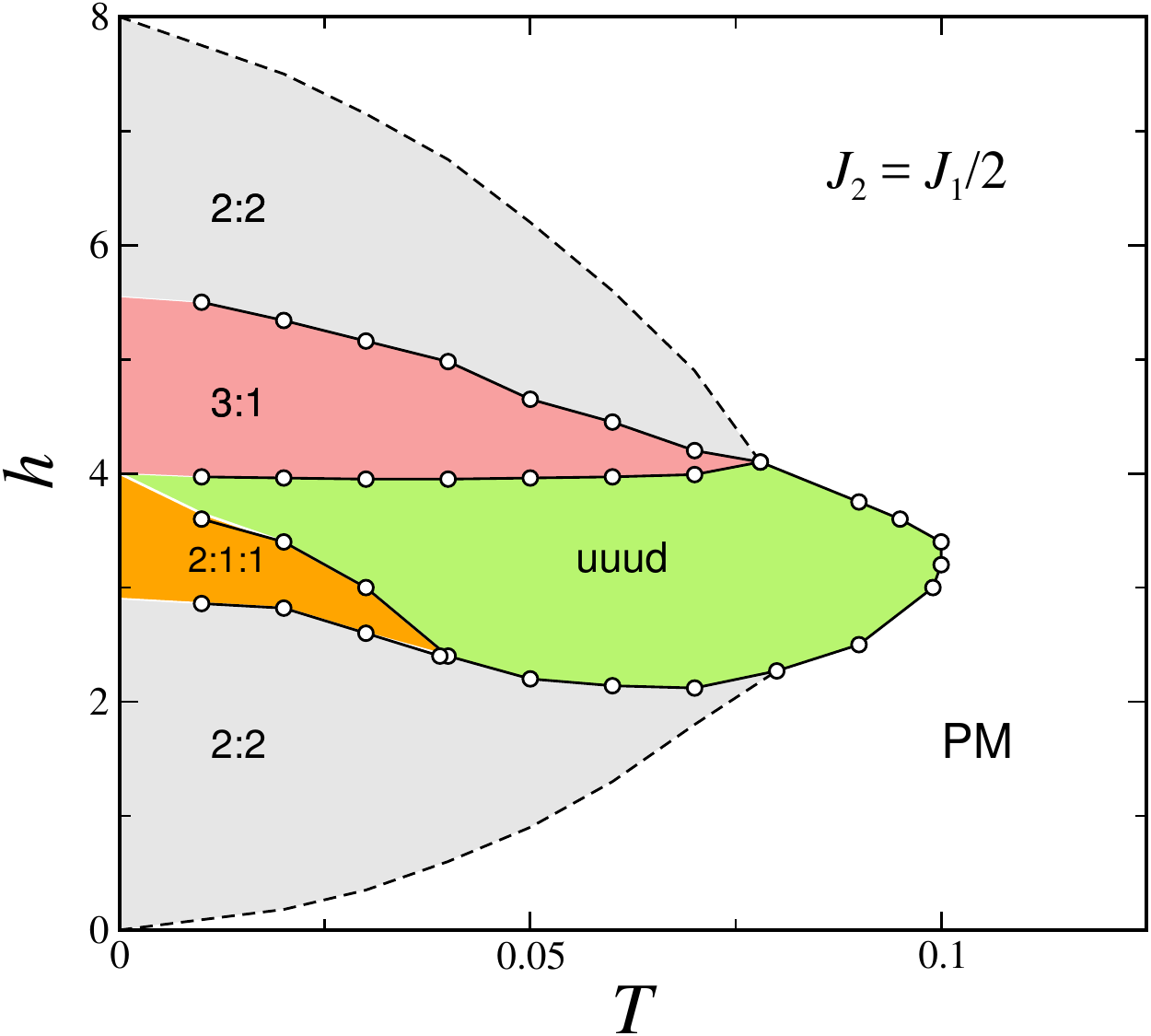}
\end{center}
\caption{Field-temperature phase diagram of the classical FSAFM  with $J_2=0.5$, $J_1=1$.
Magnetic states with (quasi) LRO are labeled according to
Fig.~\ref{SConf}, PM stands for a disordered paramagnetic state. 
Phase transition points obtained from the Monte Carlo simulations are shown by open circles
with solid lines as a guide for the eye. The dashed lines indicate the
expected Berezinski\u{\i}-Kosterlitz-Thoulesss transition  boundaries.
}
\label{Diagram}
\end{figure}

\subsection{Field-temperature phase diagram}

Next, we present new classical Monte-Carlo results on significantly larger lattices  compared to the previous studies \cite{Honecker2000a,Honecker2002}.
Simulations have been performed using the single-spin flip Metropolis algorithm in combination with the microcanonical over-relaxation moves.
Details of the algorithm are provided in recent studies
devoted to various classical frustrated spin models \cite{Zhitomirsky2008,Gvozdikova2011,Gubina2025}.

Figure~\ref{Diagram} presents the constructed $h$-$T$ phase diagram for $J_2/J_1=1/2$.  
First, we observe the expected stabilization of the $uuud$ phase by thermal fluctuations
in the regime $T \lesssim J_1/10$. The stability region is asymmetric around $h_c =0.5h_{\rm sat}$ having a significant 
extent into the low-field part of the diagram. 
Second,  there are sizable regions with the 2:1:1 and 3:1 states just below or above the $uuud$ phase, respectively.
The two noncollinear states possess the longitudinal LRO in the $s^z$ components 
accompanied by the quasi-LRO  in transverse spin components. Mapping this from magnetic to bosonic
language \cite{Matsubara1956,Matsuda1970,Huang2026}, these two phases can be identified as supersolids, whereas
the parent $uuud$ phase is an insulating state with the density-wave crystal.

\begin{figure}[t]
\begin{center}
\includegraphics[width=\columnwidth,angle=0]{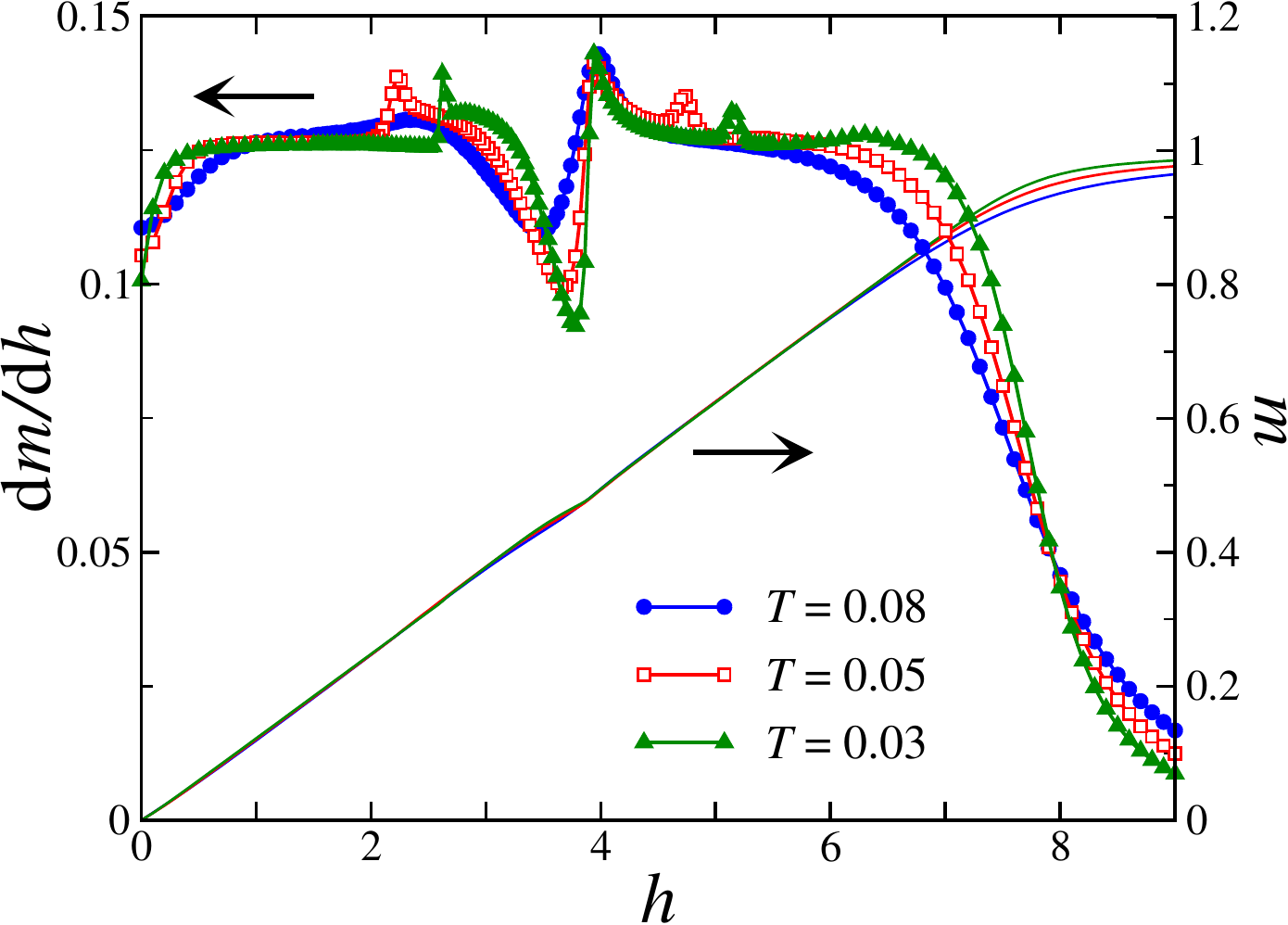}
\end{center}
\caption{The magnetization curve $m(h)$ (lines, right axis) and
field derivative of the magnetization (symbols connected by lines, left axis)
for the classical FSAFM  with $J_2=0.5$, $J_1=1$
at three different temperatures obtained on lattices with $L=32$.
}
\label{Hscan}
\end{figure}

\begin{figure}[t]
\begin{center}
\includegraphics[height=0.73\columnwidth,angle=0]{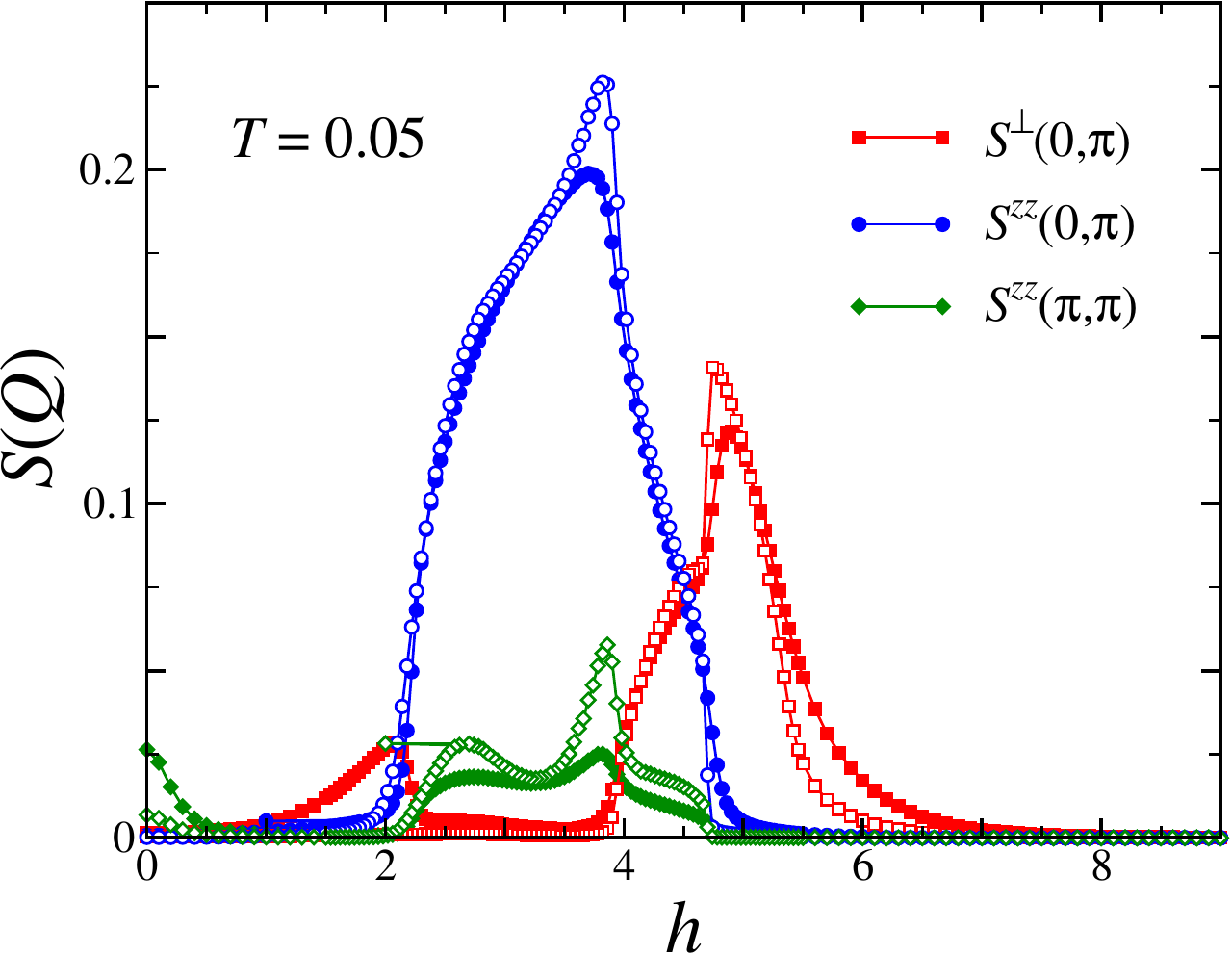}
\end{center}
\caption{The field variation of the longitudinal $S^{zz}$ and transverse $S^\perp$ spin-spin correlations
at $\bm{Q}_{\text{\Neel}}=(\pi,\pi)$ and $\bm{Q}_{\rm str}=(0,\pi)$ [summed with $(\pi,0)$] obtained for lattices with linear sizes $L=32$ (closed symbols) and
$L=64$ (open symbols).
}
\label{OPs}
\end{figure}

Figures \ref{Hscan} and \ref{OPs} illustrate some of the data used in the construction of the phase diagram Fig.~\ref{Diagram}.
According to the
analysis at the beginning of this section, the magnetization curve $m(h)$ at zero temperature is a straight line until the saturation field $h_{\rm sat}$.
Figure \ref{Hscan}  shows some magnetization curves at low but finite temperature.
The main feature that emerges at intermediate fields is a small dip for $m \lesssim 1/2$
that is most pronounced for $T/J_1=0.03$. We note that the improved Monte Carlo results
do not reproduce the hysteresis that appeared in earlier work \cite{Honecker2000a}.
To better exhibit these features in the magnetization curve, we
consider the field derivative of the magnetization ${\rm d}m/{\rm d}h$, also shown
in Fig.~\ref{Hscan}.
At zero temperature 
$m(h)$ has a constant slope
for all fields below the saturation field
$h_{\rm sat}$, {\it i.e.}, ${\rm d}m/{\rm d}h = 1/(8 J_1)$ at $T=0$ and for $h<h_{\rm sat}$.
At $T>0$, we observe in Fig.~\ref{Hscan} the emergence of small but sharp and distinct maxima
in ${\rm d}m/{\rm d}h$, which may be taken as first indications of field-induced phase transitions. For $h \lesssim 4\,J_1 = h_{\rm sat}/2$, we observe
a drop in ${\rm d}m/{\rm d}h$. The slope of the magnetization curve $m(h)$ remains finite and thus one finds only a small dip but no strict plateau in
the classical magnetization curve. Nevertheless, the reduced slope of the magnetization curve can be considered
as a distinctive feature of the $uuud$ phase.

To further characterize different phases, we look now at the static spin structure factor
$S^{\alpha\beta}(\bm{Q})$:
\begin{equation}
S^{\alpha\beta}(\bm{Q}) = \frac{1}{N^2}\: \sum_{i,j} \langle s^\alpha_i s^\beta_j\rangle \, e^{i\bm{Q}\cdot(\bm{R}_i-\bm{R}_j)} \,,
\end{equation}
where $N=L^2$ is the total number of sites and $L$ is the linear size of a simulated
lattice.
Figure~\ref{OPs} shows the Monte Carlo results for two linear lattice sizes $L=32$ and 64 to illustrate 
the role of finite-size effects.
For the considered temperature $T/J_1 = 0.05$,  the longitudinal static structure factor $S^{zz}$
indicates presence of the long-range order both for $\bm{Q}_{\text{\Neel}}=(\pi,\pi)$ and $\bm{Q}_{\rm str}=(0,\pi)$, whereas the transverse structure
factor $S^\perp$ vanishes for $2 \lesssim h/J_1 \lesssim 4$. This is exactly the behavior expected for the $uuud$ phase.
For $4 \lesssim h/J_1 \lesssim 5$, we observe the coexistence of finite amplitudes both in $S^{zz}$
and in $S^\perp(\bm{Q}_{\rm str})$, exactly as expected for the 3:1 supersolid phase.
In a striking contrast with the behavior expected from the toy  biquadratic exchange model \cite{Heinila1993,Penc2004},
the supersolid 3:1 state does not extend all the way up to the saturation at $h=h_{\rm sat}$. 
For $h/J_1 \lesssim 2$ and $h/J_1 \gtrsim 5$, Fig.~\ref{OPs} exhibits a signal
in the transverse structure factor $S^\perp(\bm{Q}_{\rm str})$, but none in the longitudinal
one $S^{zz}$. Since this corresponds to breaking of a continuous rotational symmetry
around the $z$ axis, the resulting order can be only a quasi-LRO at $T>0$ with algebraic decay of spin-spin correlations.
Transitions out of these phases are expected to be in the 
Berezinski\u{\i}-Kosterlitz-Thouless (BKT) universality class \cite{Berezinskii1971,Kosterlitz1973,Kosterlitz1974},
and indeed, the order parameter $S^\perp$ vanishes smoothly in Fig.~\ref{OPs} in the low- and
high-field limit, respectively. An accurate determination  of the BKT boundaries
 is beyond the scope of  the present study. Instead, we indicate their presence on the phase diagram
by grey regions in Fig.~\ref{Diagram}.

\begin{figure}[t!]
\begin{center}
\includegraphics[height=0.73\columnwidth,angle=0]{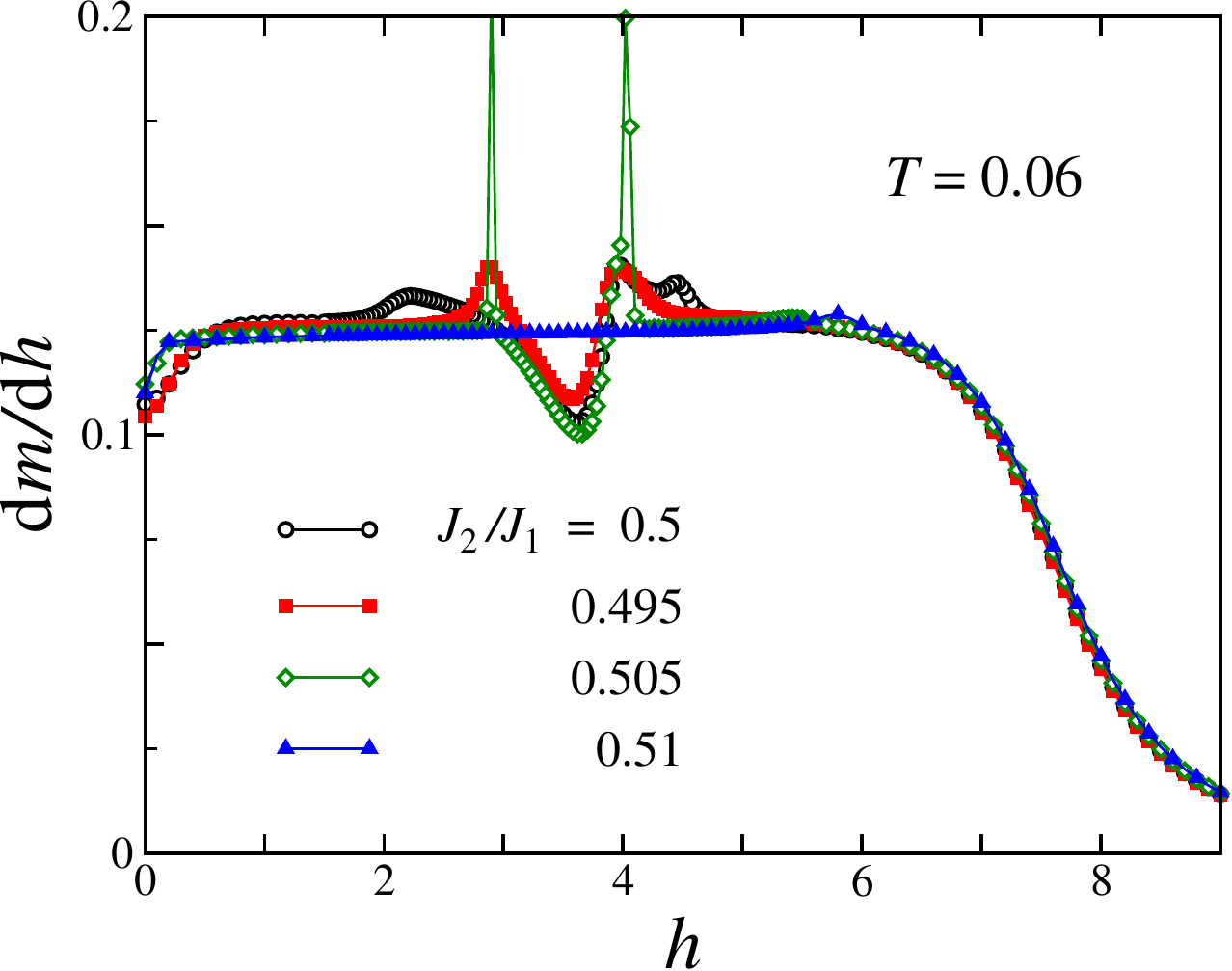}
\end{center}
\caption{Field derivative of magnetization for the classical FSAFM
at  $T= 0.06$ for different exchange ratios close
to  $J_2=0.5$ (in units of $J_1=1$).}
\label{fig:Hchi_T_0.06comp}
\end{figure}

\begin{figure}[t!]
\begin{center}
\includegraphics[height=0.75\columnwidth,angle=0]{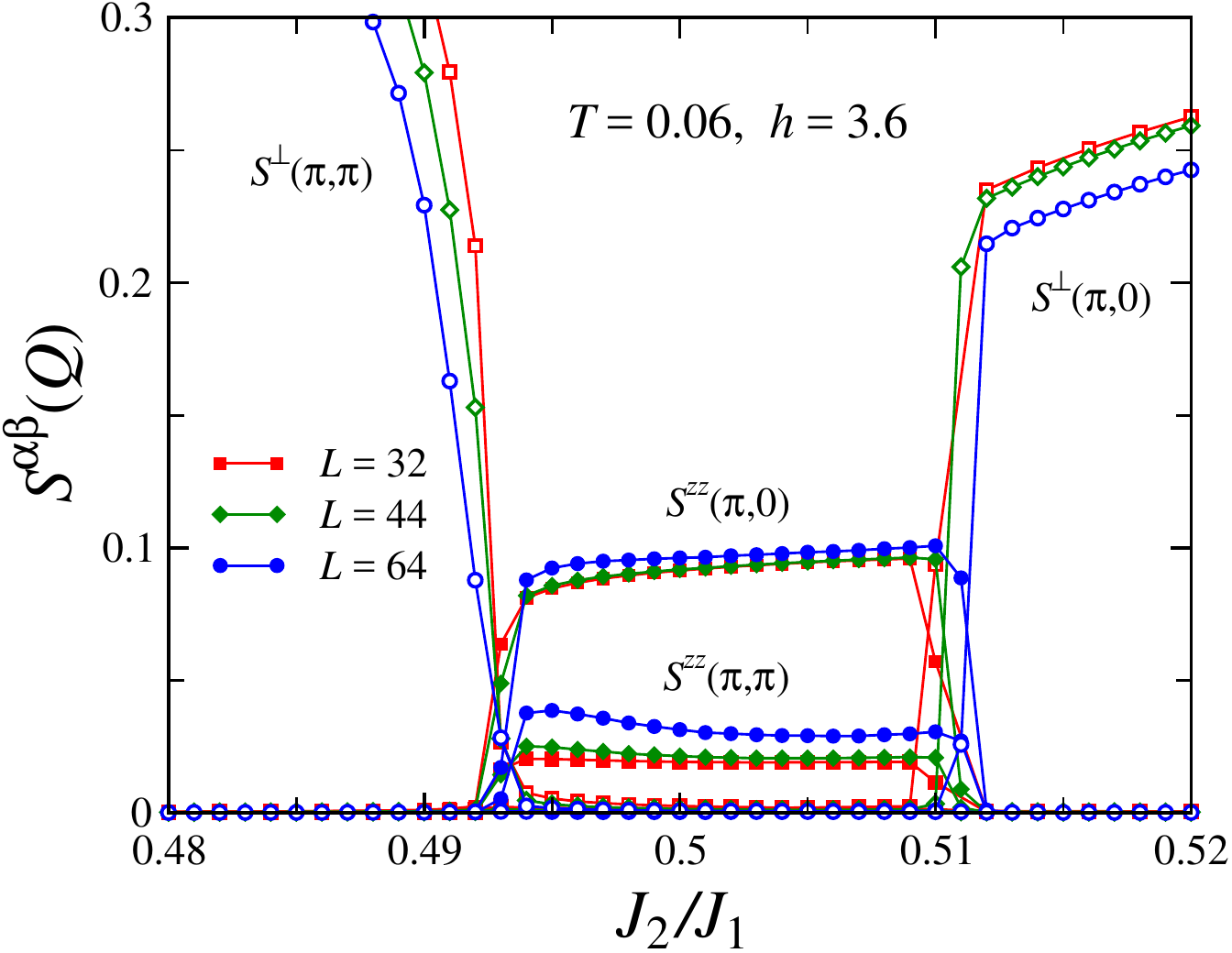}
\end{center}
\caption{Static structure factor of the classical FSAFM for fixed
$T=0.06$ and $h=3.6$ (in units of $J_1=1$) as a function of $J_2/J_1$ for three lattice sizes $L$.}
\label{fig:Fig5bis}
\end{figure}

Figures \ref{fig:Hchi_T_0.06comp} and \ref{fig:Fig5bis} illustrate stability of the $uuud$ state  
upon moving away from the strongly frustrated point $J_2 \neq 0.5J_1$. Focusing first on field scans in Fig.~\ref{fig:Hchi_T_0.06comp} 
for $T=0.06 J_1$, one can clearly observe sharp anomalies in the field derivative of the magnetization 
${\rm d}m/{\rm d}h$ for $J_2/J_1 = 0.5 \pm 0.005$ indicating the upper and lower boundaries of the
the $uuud$ plateau state. In contrast, 
${\rm d}m/{\rm d}h$ remains completely flat for $J_2/J_1 = 0.51$ featuring canted \Neel\ and stripe states in the entire range of
magnetic fields. For $T=0.06 J_1$ and $h=3.6 J_1$,  Fig.~\ref{fig:Fig5bis} shows the dependence of the transverse and longitudinal 
components  of the static structure factor upon varying the ratio of $J_2/J_1$. Here, one can see the same trend with the $uuud$ state
clearly stabilized by thermal fluctuations in a finite range of $J_2/J_1$.
Note that the $uuud$ state is the classical ground state only for a single frustration ratio $J_2/J_1 = 1/2$. Hence,
for $J_2/J_1 \neq 1/2$, the region of the $uuud$ phase is detached from the vertical axis of the phase diagram at $T=0$ and requires
sufficiently strong thermal fluctuations for its stabilization. Conversely, for $J_2/J_1=1/2$ the strong thermal 
fluctuations for $T \gtrsim 0.1J_1$ destroy the $uuud$ LRO, see Fig.~\ref{Diagram}. Thus, one requires a delicate balance of sufficiently
but not too strong thermal fluctuations to stabilize the $uuud$ state, consistent with its disappearance
when $J_2/J_1$ deviates by about $0.01$ from the degenerate ratio $1/2$.

\section{Quantum model}
\label{sec:Quantum}

We start with a basic consideration that is valid for any spin quantum number $s$: the single-magnon dispersion above the ferromagnetically polarized
background is a single-particle problem that is straightforward to solve. One finds the single-magnon dispersion to be given by (compare also Ref.~\cite{Honecker2000b})
\begin{eqnarray}
\omega(k_x,k_y) &=& 2 s \bigl(J_1 \left( \cos{k_x} + \cos{k_y} \right) \nonumber \\
&& \qquad +  J_2 \left( \cos(k_x+k_y) + \cos(k_x-k_y) \right) \bigr) \nonumber \\
&& - 4 s \left(J_1 + J_2\right) + h \nonumber \\
&=& 2 s \left(J_1 \left( \cos{k_x} + \cos{k_y} \right)
+  2 J_2 \cos{k_x} \cos{k_y} \right) \nonumber \\
&& - 4 s \left(J_1 + J_2\right) + h \, . \label{omegaK}
\end{eqnarray}
If the transition to saturation is a second-order quantum phase transition, the minimal single-magnon energy
vanishes exactly at the saturation field $\omega(\bm{k}_{\min}) - h_{\rm sat} = 0$. Applying this
condition to Eq.~(\ref{omegaK}), we can determine the saturation field as
\begin{equation}
h_{\rm sat} = \begin{cases}
8 s J_1 & \text{for } J_2 \le J_1/2 , \\
4 s (J_1+2J_2) & \text{for } J_2 \ge J_1/2 .
\end{cases}
\label{eq:hsat}
\end{equation}
In particular $h_{\rm sat} = 8 s J_1$ at $J_1=J_2/2$.
In the classical limit $s \to \infty$ and upon suitably rescaling
the spins by $1/s$, {\it i.e.}, by effectively replacing $s \to 1$,
we find $h_{\rm sat} = 8 J_1$ at $J_1=J_2/2$ in the classical limit,
as already stated in Subsec.~\ref{sec:ClassGS}.

An interesting special case is $J_2=J_1/2$ where the dispersion relation (\ref{omegaK})
simplifies (see also Ref.~\cite{Jackeli04})
\begin{eqnarray}
\left.\omega(k_x,k_y)\right|_{J_2=J_1/2}
 &=& 2sJ_1 \left(\left(1 + \cos k_x\right) \left(1+\cos k_y\right) -4\right) \nonumber \\
 && + h
\label{omegaKhalfG}
\end{eqnarray}
such that
\begin{equation}
\left.\omega(k_x,\pi)\right|_{J_2=J_1/2}
 = \left.\omega(\pi,k_y)\right|_{J_2=J_1/2}
 = - 8 s J_1 + h \, ,
\label{omegaKhalf}
\end{equation}
{\it i.e.}, we find degenerate lines of minima for $h \ge h_{\rm sat}$.
This is reminiscent of the classical degeneracy discussed in Subsec.~\ref{sec:ClassGS},
in particular the degenerate manifold evolving from the $uuud$ state at $m=1/2$.

\begin{figure}[t!]
\begin{center}
\includegraphics[width=0.95\columnwidth]{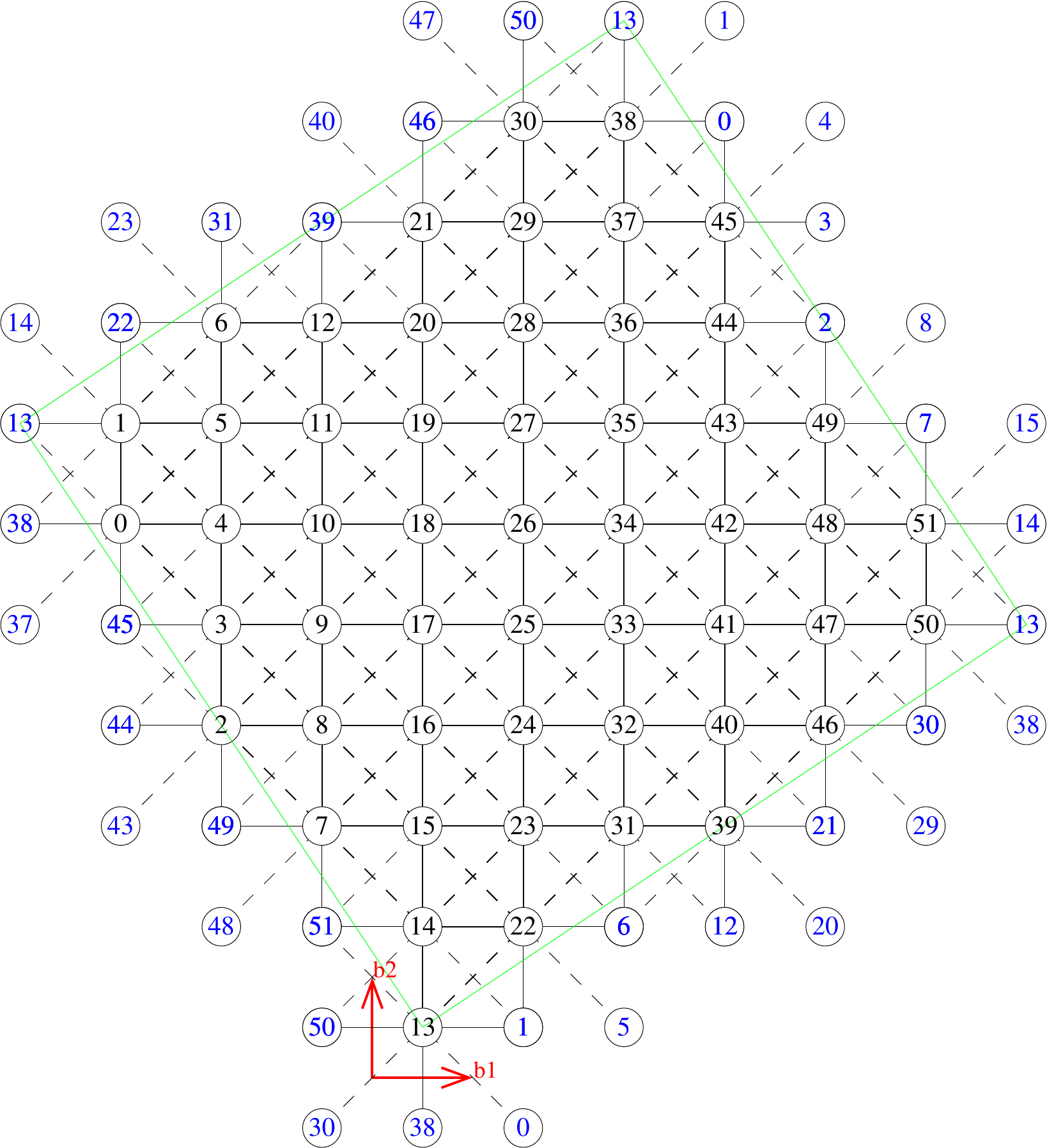}
\end{center}
\caption{The $J_1$-$J_2$ model on a finite square lattice of $N=52$ sites.
Repeated sites (shown in blue) indicate periodic boundary conditions.
}
\label{fig1}
\end{figure}

\subsection{Zero-temperature $s=1/2$ phase diagram}

\label{sec:qGS}

To numerically investigate the quantum case at finite magnetization $m<1$, we now restrict
to the extreme quantum limit $s=1/2$. We employ the Lanczos method \cite{Lanczos1950,Cullum2002}
\begin{figure*}[ht!]
\begin{center}
\includegraphics[width=0.8\columnwidth]{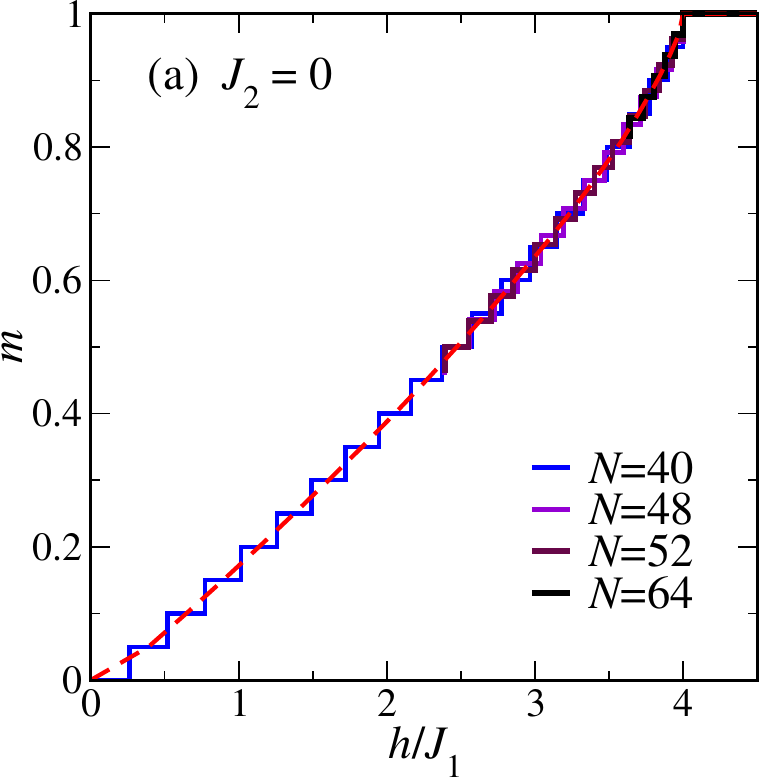}\qquad%
\includegraphics[width=0.8\columnwidth]{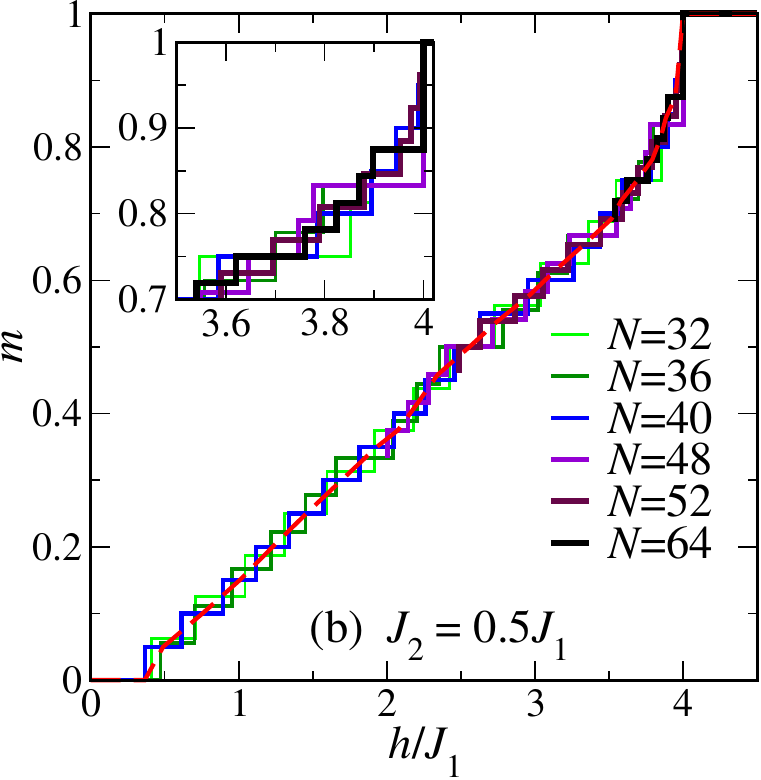} \\
\includegraphics[width=0.8\columnwidth]{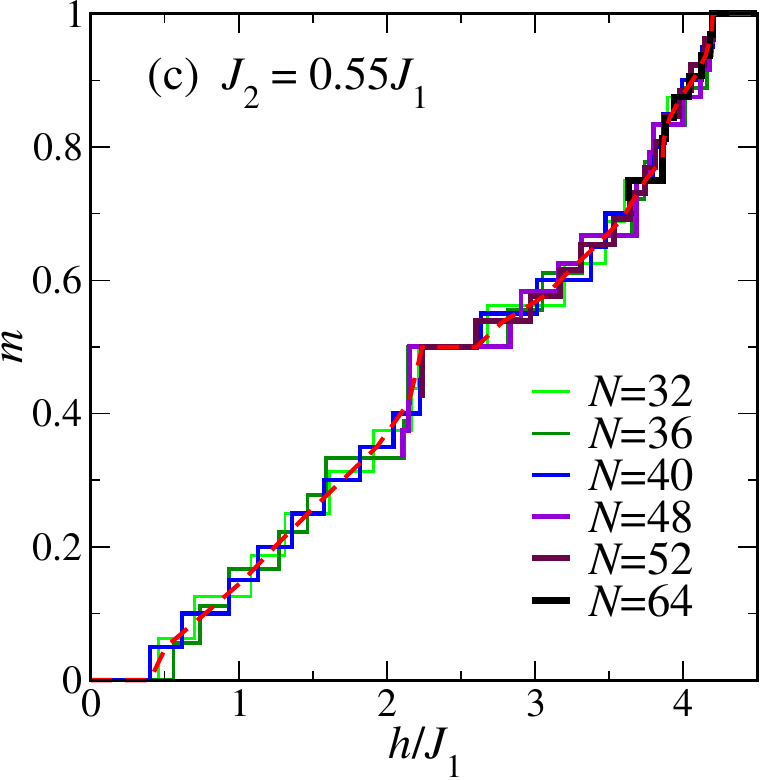}\qquad%
\includegraphics[width=0.8\columnwidth]{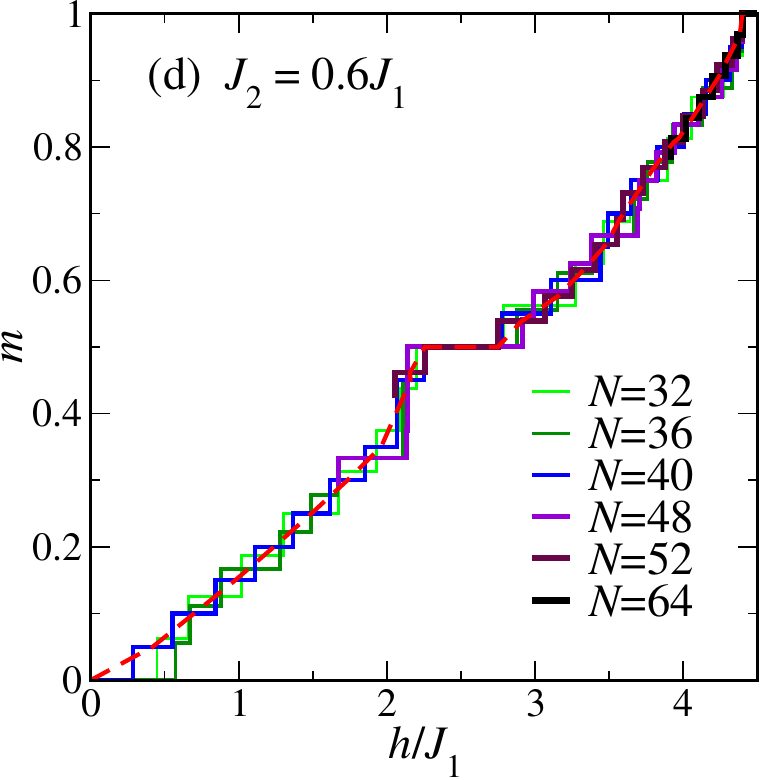}
\end{center}
\caption{Zero-temperature magnetization curves $m(h)$ of the spin-1/2 FSAFM for 
$J_2/J_1=0$ (a), $0.5$ (b), $0.55$ (c), and $0.6$ (d).
Full curves denote finite-size results for the given system size $N$.
The dashed red lines indicate an estimate for the behavior of the infinite system.
This estimate is obtained by connecting the midpoints of the finite-size steps
for the largest available system size $N$ \cite{Bonner1964}, except for suspected plateaux (including the transition
to saturation $m=1$), where we use the corners of the steps instead.
For a discussion why we draw $m=1/2$ plateaux in the extrapolated curves at $J_2/J_1=0.55$ and $0.6$, but none for $J_2/J_1=0.5$, compare Fig.~\ref{fig:plateau} below and the discussion in the text;
for the identification of this $m=1/2$ plateau with the $uuud$ state, see
Fig.~\ref{fig:Cor} below and the discussion in the text.
}
\label{fig:M}
\end{figure*}
to find the minimal eigenvalues and -vectors of the Hamiltonian and thus determine its
ground-state properties. We use in particular $S^z$ conservation. The total Hilbert
space for a given $S^z$ and $N$ spins 1/2 has dimension ${N \choose N/2 - S^z}$. Consequently,
the dimension is maximal for $S^z = 0$ and decreases with increasing $S^z$, {\it i.e.},
higher magnetic fields. In particular, we can push computations to the $N=52$ lattice
shown in Fig.~\ref{fig1} for $m \gtrsim 1/2$, {\it i.e.}, $S^z \gtrsim N/4 = 13$.
This covers in particular the region relevant to a possible $m=1/2$ plateau.
Note that this $N=52$ lattice shares the $90^\circ$ rotational symmetry of the infinite lattice,
but it breaks spatial reflection symmetries.

\subsubsection{Magnetization curves}

Figure~\ref{fig:M} presents zero-temperature magnetization curves, where we normalized the
saturation magnetization to $m=1$.
In addition to finite-size results, we show estimates for the behavior of the infinite system
by the dashed red lines. Following Ref.~\cite{Bonner1964},
this estimate is in general obtained by connecting the midpoints of the finite-size steps
for the largest available system size $N$. The exception are suspected plateaux: here we use the corners of the steps instead. The main bias in this extrapolated curves is whether
or not we expect a plateau and we refer to the discussion below for further details on this point.

For reference, we include the plain square-lattice case
with $J_2=0$ (Fig.~\ref{fig:M}(a)) that has been investigated in detail previously,
see for example Ref.~\cite{Honecker2004}, including the $N=40$ results.
Here, we find a smooth curve in the thermodynamic limit with an upwards curvature
that ends in a logarithmic singularity. This behavior is characteristic of a two-dimensional
quantum spin system, and has been analytically predicted for the present case
by second-order spin-wave theory \cite{Nikuni1998}.

Figure~\ref{fig:M}(b) presents the results for the highly frustrated case $J_2/J_1=1/2$.
Here, we have no evidence for particular features around $m=1/2$ and thus have drawn a smooth
curve as estimate for the behavior in the limit $N \to \infty$.
However, the transition to saturation is exceptionally steep in this case,
as is demonstrated by the inset to Fig.~\ref{fig:M}(b). This behavior can be
traced to the degeneracy of the one-magnon spectrum arising at $J_2/J_1=1/2$,
see Eq.~(\ref{omegaKhalf}). This degeneracy gives rise to excitations
localized along horizontal or vertical lines across the lattice \cite{Schulenburg2002}
and thus leads to a finite-size jump to
saturation of size $\delta m = 1/L$ for an $L \times L$ square geometry.
In the thermodynamic limit, this is expected to give rise to a square-root singularity
with a logarithmic correction at the saturation field $h_{\rm sat}$
\cite{Jackeli04}.

Next, we consider the case $J_2/J_1=0.55$, see Fig.~\ref{fig:M}(c). Here, the $N=48$
system appears to be an outlier, but the other system sizes converge to a stable $m=1/2$ plateau.
The resulting magnetization curve is in good agreement with results from the
density matrix renormalization group (DMRG) method \cite{Shibata2016}, and even
the estimates of the lower and upper critical fields of the $m=1/2$ plateau are
in reasonable quantitative agreement. Note furthermore that the slope of the magnetization curve
is very steep when it approaches $m=1/2$ from below. This might be a signature of a first-order
quantum phase transition, although the available data does not suffice to draw a definite
conclusion on the nature of this field-induced phase transition.

Finally, we look at $J_2/J_1=0.6$. The resulting magnetization curve, shown in
Fig.~\ref{fig:M}(d), is qualitatively similar to the case $J_2/J_1=0.55$. In particular,
we find again an $m=1/2$ plateau that might be even broader than in the previous case.
Our results are again consistent with the DMRG ones \cite{Shibata2016} for $J_2/J_1=0.6$.

We note that a Chern-Simons theory predicted a distinct $m=1/3$ plateau in addition to the
$m=1/2$ one \cite{Chern2002}. The $N=36$ and $48$ systems indeed exhibit a broader step for
$J_2/J_1=0.55$ (Fig.~\ref{fig:M}(c)) and $0.6$ (Fig.~\ref{fig:M}(d)) while
the magnetization curves for the other system sizes appear to exhibit a smooth
behavior in this region.
The appearance of a strict $m=1/3$ plateau is in fact only possible for system
sizes divisible by three. It is difficult to say on the basis of the present
data if sizes $N$ that are multiplies of three favor a finite-size plateau
or if sizes not commensurate with three miss a one-third plateau.
We therefore chose to follow Ref.~\cite{Shibata2016} and
have not drawn an $m=1/3$ plateau in our extrapolated curves.

Lastly, we note that at $J_2/J_1=0.5$ and $J_2/J_1=0.55$ we have drawn an $m=0$ plateau
in Fig.~\ref{fig:M}(b) and (c), respectively. This corresponds to a spin gap in zero
field, as is to be expected for the paramagnetic phases around $J_2/J_1=1/2$
mentioned in the Introduction. The situation for $J_2/J_1=0.6$ (Fig.~\ref{fig:M}(d))
is less clear. Here, the $N=36$ system suggests the presence of a spin gap, but this is
not confirmed by the $N=40$ system. We therefore have not drawn an $m=0$ plateau although
we cannot exclude the existence of a small $m=0$ plateau, {\it i.e.}, spin gap for
$J_2/J_1=0.6$. In any case, we recall that the zero-field behavior is a challenging 
issue that is beyond the scope of the present work.

\begin{figure}[t!]
\begin{center}
\includegraphics[width=0.8\linewidth]{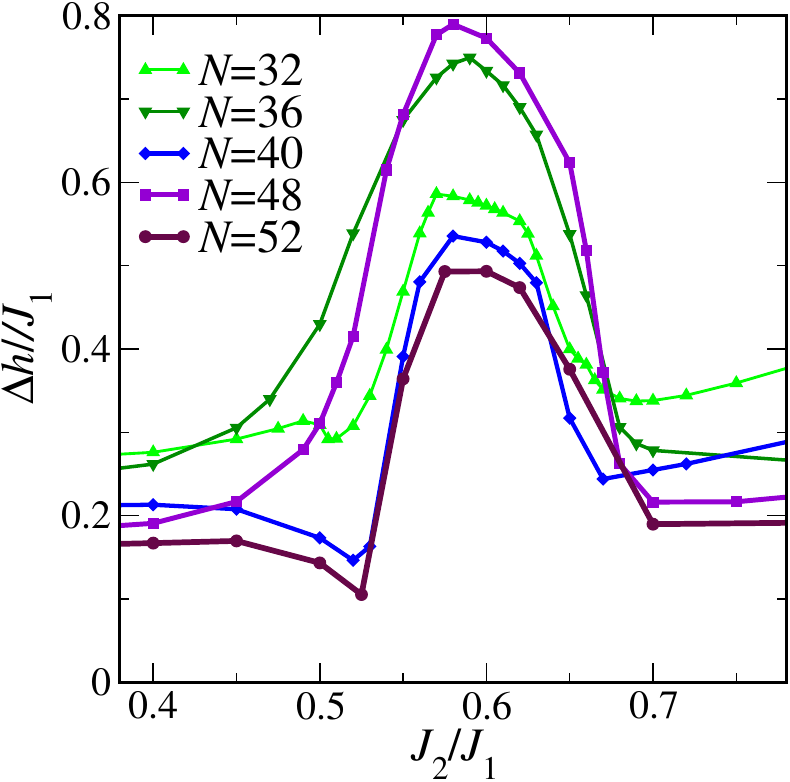}
\end{center}
\caption{Width $\Delta h$ of the $m=1/2$ step in the magnetization curve as a function of the frustration
parameter $J_2/J_1$.
Symbols represent actual data points whereas lines are guides to the eye.
}
\label{fig:plateau}
\end{figure}

\begin{figure}[t!]
\begin{center}
\includegraphics[width=0.8\columnwidth]{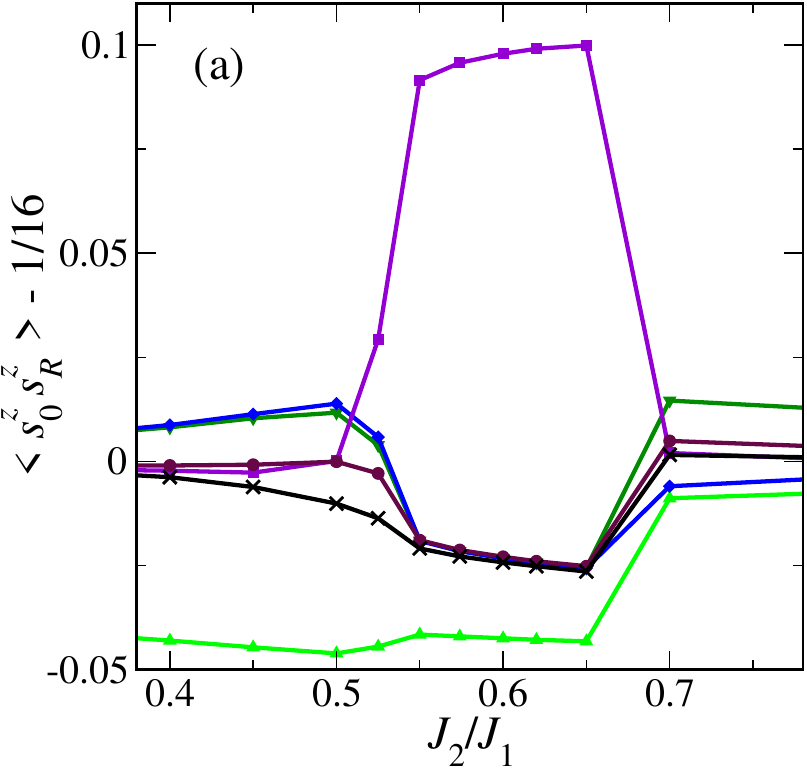}\\
\includegraphics[width=0.8\columnwidth]{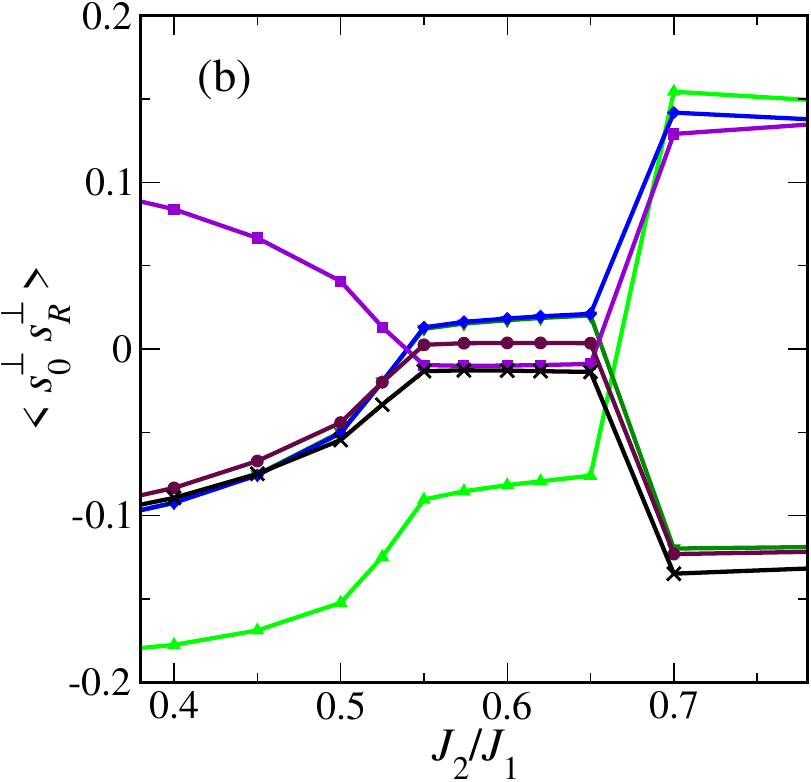} 
\end{center}
\caption{Real-space correlations on the $N=52$ lattice for $m=1/2$ ($S^z = 13$)
as a function of $J_2/J_1$.
Panel (a) shows the  connected logitudinal correlations, panel
(b) the transverse ones\footnote{%
Reliable information about the corresponding values of $R$
is no longer available, as neither is the relevant author. We therefore refrain from labeling the
curves and mention just that the same symbols and colors are used for transverse and longitudinal
correlations at the same $R$.
}.
Symbols represent actual data points whereas lines are guides to the eye.
}
\label{fig:Cor}
\end{figure}

\subsubsection{$m=1/2$ plateau}

\label{sec:m1o2}

Now we focus on a more detailed characterization of the $m=1/2$ plateau.
First, we look at the width of the $m=1/2$ step, calculated
as $\Delta h = E(N/4+1)-2E(N/4)+ E(N/4-1)$, where $E(S^z)$ is the
energy of the ground state for the given value of $S^z$ at $h=0$.
Figure~\ref{fig:plateau} presents the behavior of $\Delta h$ as a function of $J_2/J_1$
for systems with $32 \le N \le 52$ spins 1/2.
We have chosen lattices with an aspect ratio as close as possible to unity
for a given size $N$, but nevertheless the results
fall into two groups, namely the square respectively rectangular geometries for $N=36$ and $48$ on the one side,
and the tilted cases $N=32$, $40$, and $52$ on the other side. 
This non-monotonic finite-size behavior renders an extrapolation to $N\to \infty$
difficult. Nevertheless, we observe a clear and consistent enhancement of the $m=1/2$ step size in the region
$0.52 \lesssim J_2/J_1 \lesssim 0.7$, which we thus take as a first estimate for the stability
region of the $m=1/2$ plateau.

To gain further insight into the state of the $m=1/2$ plateau, Fig.~\ref{fig:Cor} shows real-space
spin-spin correlations at distance $R$ for the $N=52$ lattice of Fig.~\ref{fig1}, {\it i.e.},
at $S^z=13$.
Note that $\langle s^z_R \rangle = m/2 = 1/4$.
We therefore subtract a background of $\langle s^z_0\rangle \langle s^z_R\rangle =1/16$
from the longitudinal correlation function  $\langle s^z_0 s^z_R\rangle$ to obtain
the connected one, that is shown in Fig.~\ref{fig:Cor}(a). The transverse one
in Fig.~\ref{fig:Cor}(b) is obtained as
$\langle s^\perp_0 s^\perp_R\rangle =
\langle \bm{s}_0 \cdot \bm{s}_R\rangle -
\langle s^z_0 s^z_R\rangle$.
Overall, we observe three distinct regimes in  Fig.~\ref{fig:Cor}:
(i) For $J_2/J_1 \lesssim 0.5$ the transverse correlations dominate.
This is consistent with {\Neel} order in the transverse components, as expected in the
limit $J_2 = 0$.
(ii) For $0.52 \lesssim J_2/J_1 < 0.7$, the transverse correlations essentially vanish,
but strong longitudinal ones develop. This is consistent with the collinear $uuud$ state that is
expected for an $m=1/2$ plateau from the investigation of the classical model in Sec.~\ref{sec:Class}.
(iii) For $J_2/J_1 > 0.7$, the longitudinal components vanish again while a new
signal develops in the transverse components. Different signs reflect an order that differs
from the one in regime (i). Indeed, for $J_1=0$, one has two interpenetrating square lattices,
each of which develops independent {\Neel} order. A weak coupling $J_1 > 0$ then
leads to a ``stripe'' phase, consistent with the correlations observed in Fig.~\ref{fig:Cor}(b)
for $J_2/J_1 > 0.7$.

Taken together, the data for $\Delta h$ and the spin-spin correlation functions at $m=1/2$ are consistent
with an $uuud$ plateau in the regime $0.52 \lesssim J_2/J_1 \lesssim 0.7$,
gapless transverse {\Neel} order for $J_2/J_1 \lesssim 0.52$,
and gapless transverse stripe order for $J_2/J_1 \gtrsim 0.7$.
These results on the stability of the $uuud$ state for the $s=1/2$ model are in sharp contrast with the
thermal order by disorder for classical spins obtained above in Sec.~\ref{sec:Class}. Specifically, for a fixed $T$
the $uuud$ state appears rather symmetrically for $0.492 \lesssim J_2/J_1 \lesssim 0.51$ in the classical model, see
Fig.~\ref{fig:Fig5bis}. In fact, this behavior 
illustrates a remarkable distinction between quantum and thermal order-by-disorder effects already mentioned in the end of Sec.~\ref{sec:ClassGS}.
Using a so-called augmented spin-wave theory for large spins, Coletta {\it et al.}~\cite{Coletta2013} have demonstrated 
such an asymmetric stabilization of the 1/2 plateau for $J_2/J_1\gtrsim 0.5$. This effect is driven by the high-energy magnons
with $\omega(\bm{k})\approx Js$, which contribute to the vacuum zero-point energy on the same footing with the low-energy
magnons. On the other hand, the thermal order by disorder is completely dominated by the low-energy magnons with $\omega(\bm{k}) \approx T$.

\subsection{Thermodynamics for $s=1/2$}

\label{sec:qT}

Finally, we investigate the finite-temperature behavior of the $s=1/2$ FSAFM.
This is motivated by two questions. Firstly, we are interested in the stability of the $uuud$
state at finite temperature and secondly we intend to verify the impact of the competing
interactions on the magnetocaloric effect, in particular around the degeneracy arising
at the saturation field $h_{\rm sat}$ for $J_2/J_1 = 1/2$.
To address these questions, we focus on the two cases $J_2/J_1 =  0.5$ and $0.55$.

In order to obtain results for system sizes $N$ that are comparable to those that we have
investigated at $T=0$, we employ the finite-temperature Lanczos method
\cite{JaP:PRB94,JaP:AP00,Hams2000,ScW:EPJB10,prelovsek_bonca_2013,HaS:EPJB14,PRE:COR17,kago42,Wietek2019,Accuracy_FTL_PRR2020,Seki_FTLM2020}.
To be precise, we have employed two different implementations of this
method. A first one follows the historic route
\cite{JaP:PRB94,JaP:AP00,ScW:EPJB10,prelovsek_bonca_2013,HaS:EPJB14,PRE:COR17,kago42}
whereas a second one builds on the notion of ``thermal pure quantum states'' (TPQ)
\cite{Sugiura2012,Sugiura2013}.
With the Krylov approximation to the imaginary-time 
evolution~\cite{Hochbruck1997}, the two methods become equivalent \cite{Honecker2020},
but the technical implementations are still different, and the second point of
view lends a more solid theoretical foundation to the earlier heuristic approach.
In order to distinguish the two methods, we refer to the historical approach as ``FTLM''
and to the more recent implementation \cite{Wietek2019} as ``TPQ''.

\begin{figure}[t!]
\begin{center}
\includegraphics[width=0.8\columnwidth]{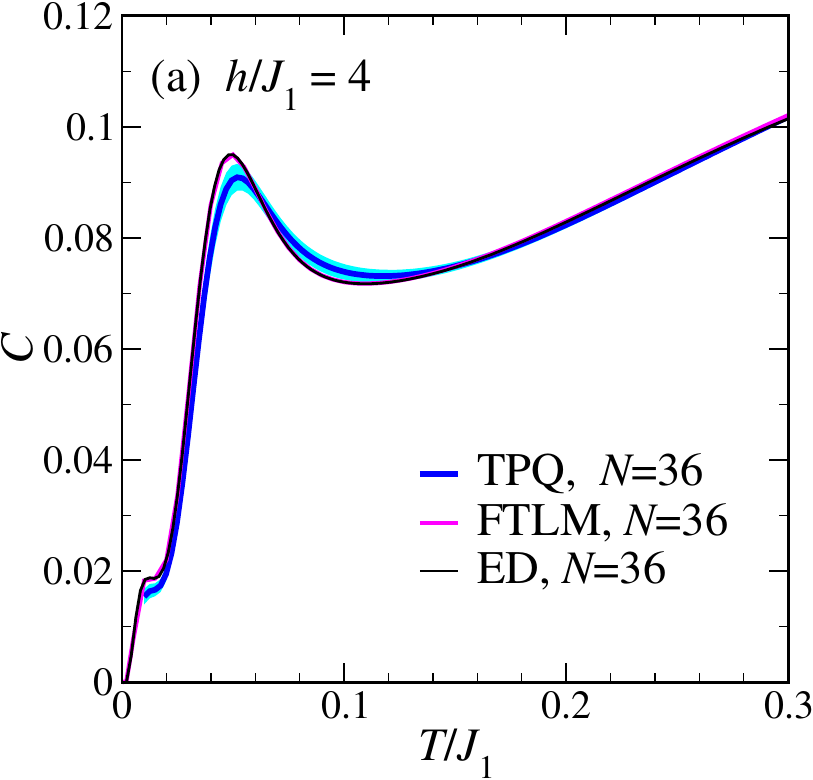}\\
\includegraphics[width=0.8\columnwidth]{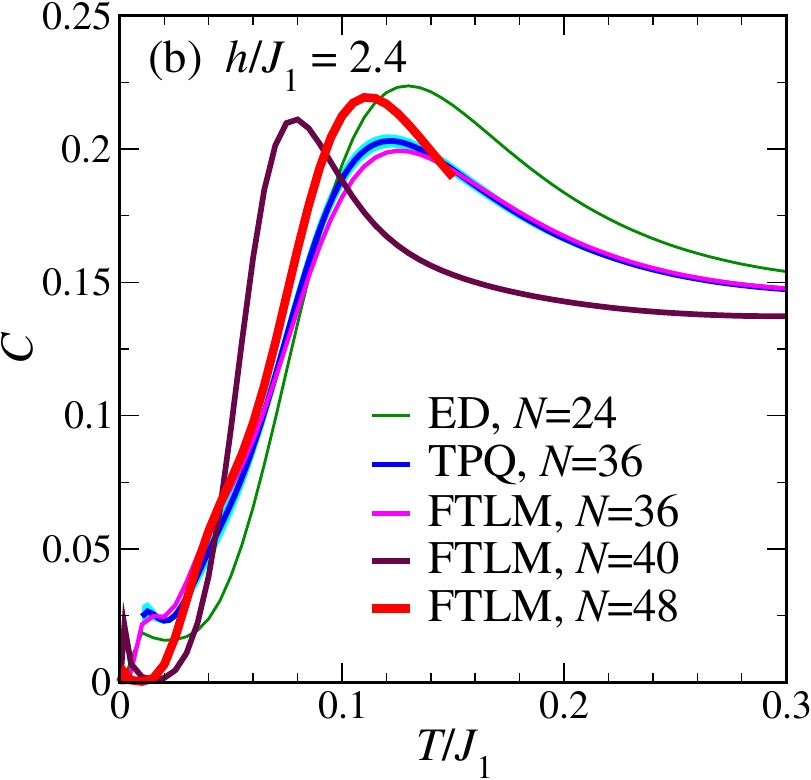} 
\end{center}
\caption{Specific heat per spin of the $s=1/2$ FSAFM at $J_2/J_1 = 0.55$
for $h/J_1=4$ (a) and $h/J_1=2.4$ (b).
Panel (a) compares three different methods for the $N=36$ system.
Low-temperature features likely reflect the finite size of the system in this case.
Panel (b) presents results for four different system sizes.
In this case, features for $T \lesssim 0.02 J_1$ are likely artifacts of either
the finite size or the computational method whereas the peak for $T \approx 0.1 J_1$
is taken as a signature of the phase transition into the $m=1/2$ $uuud$ state.
}
\label{fig:Ch}
\end{figure}

\subsubsection{Specific heat}

Figure~\ref{fig:Ch} presents the specific heat per spin $C$ as a function of temperature at $J_2/J_1 = 0.55$ for two selected fields $h/J_1 = 4$ and $2.4$.
TPQ and FTLM are subject to statistical errors that arise from sampling of the
start vectors, corresponding to the $T=\infty$ limit.
The cyan error tube around the $N=36$ TPQ result has been obtained with a
jackknife analysis \cite{Efron1981}, whereas no error bars are shown for the FTLM
results in Fig.~\ref{fig:Ch}.

The first case $h/J_1 = 4$ lies just below the saturation field that is $h_{\rm sat}/J_1 = 4.2$ for $J_2/J_1 = 0.55$ (see Eq.~(\ref{eq:hsat})). The result for the $N=36$ system is shown in Fig.~\ref{fig:Ch}(a)
and serves as a benchmark.
The proximity to saturation permits us to perform exact diagonalization (ED) that covers the complete spectrum up to sufficiently high energies to yield essentially exact results for the entire temperature range shown. This ED curve is free from the statistical errors that are inherent to FTLM and thus serves as a reference. We see that the FTLM curve is very close to the ED one, and also the TPQ curve deviates from these two only by an amount that is consistent with the estimated error.
The features observed in Fig.~\ref{fig:Ch}(a) around $T/J_1 = 0.01$ and $0.05$ are most likely due to finite-size effects reflecting the finite system size $N=36$ and thus have no physical meaning.

Figure~\ref{fig:Ch}(b) shows results for $h/J_1=2.4$. This value of the field is chosen to lie in the middle of the $m=1/2$
plateau of the $T=0$ magnetization curve, see Fig.~\ref{fig:M}(c).
TPQ and FTLM for the $N=36$ system are again consistent with each other. Beyond this, here we include
results for other system sizes, namely $N=24$ ED and also FTLM for $N=40$ and $48$. We note that for $N\le 40$ we have sampled all $S^z$ sectors,
but for $N=48$ this is no longer possible. Validity of the $N=48$ FTLM results is thus restricted to high magnetic fields and low temperatures.
In particular for $h/J_1=2.4$, we have to restrict the $N=48$ FTLM curve to $T/J_1 \le 0.15$.
Nevertheless, all curves in Fig.~\ref{fig:Ch}(b) exhibit a maximum in $C(T)$ around
$T/J_1 \approx 0.1$.
The system sizes studied here are larger than those of previous investigations \cite{Shannon2004,Schmidt2005,Schmidt06,Schmidt07a,Schmidt07b,Seabra2009,Schmidt2017},
but they are still too small to permit an unambiguous determination of a thermodynamic phase transition.
Nevertheless, the $uuud$ state breaks translational symmetry such that we expect a finite-temperature phase transition out of
it, as in the classical version of the model. The maxima of the specific heat in  Fig.~\ref{fig:Ch}(b) may serve as an estimate
for the location of the phase transition such that we infer a transition temperature of the order
of $T=J_1/10$ for  $h/J_1=2.4$.
By contrast, the upturns observed at low temperature in the $N=40$ and $48$ FTLM curves are likely to be artifacts of the method,
while the true low-temperature symptotics of $C(T)$ is expected to be an exponential decrease for $T\to0$ on any system of a finite size $N$.

\begin{figure*}[ht!]
\begin{center}
\includegraphics[width=0.95\columnwidth]{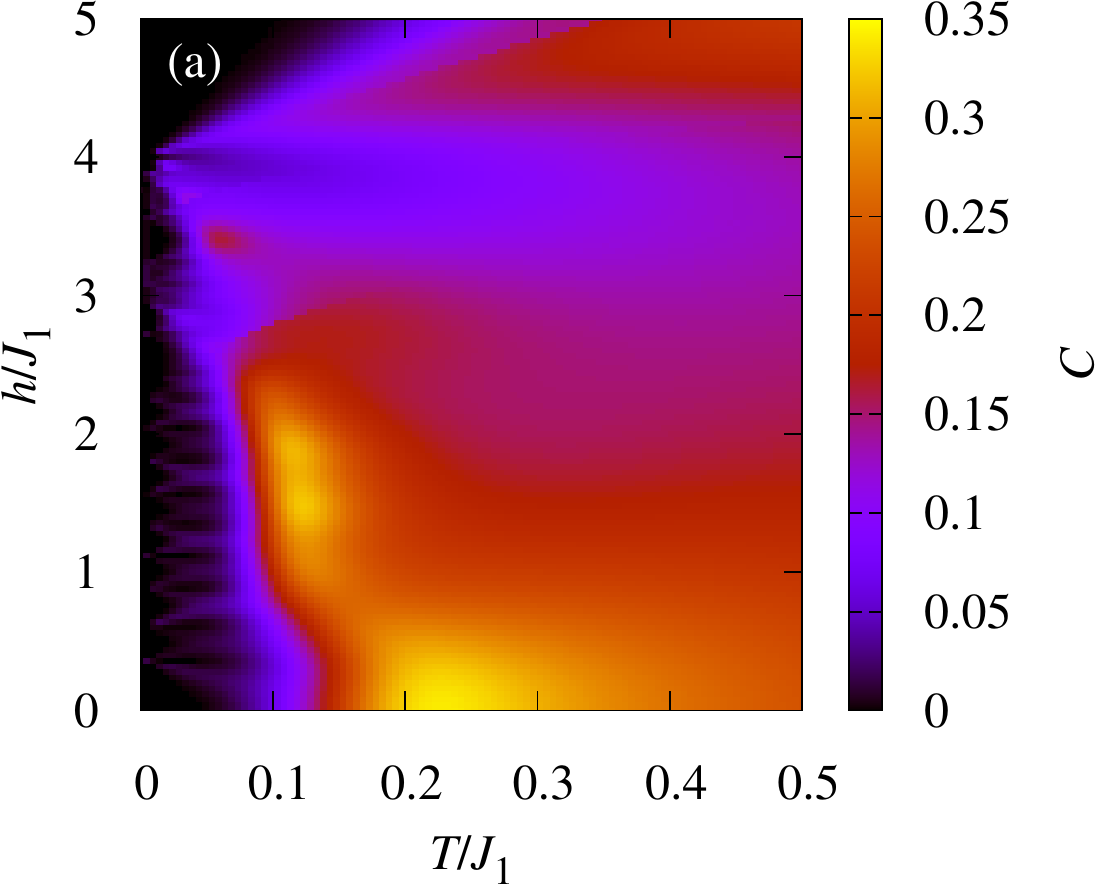}\qquad%
\includegraphics[width=0.95\columnwidth]{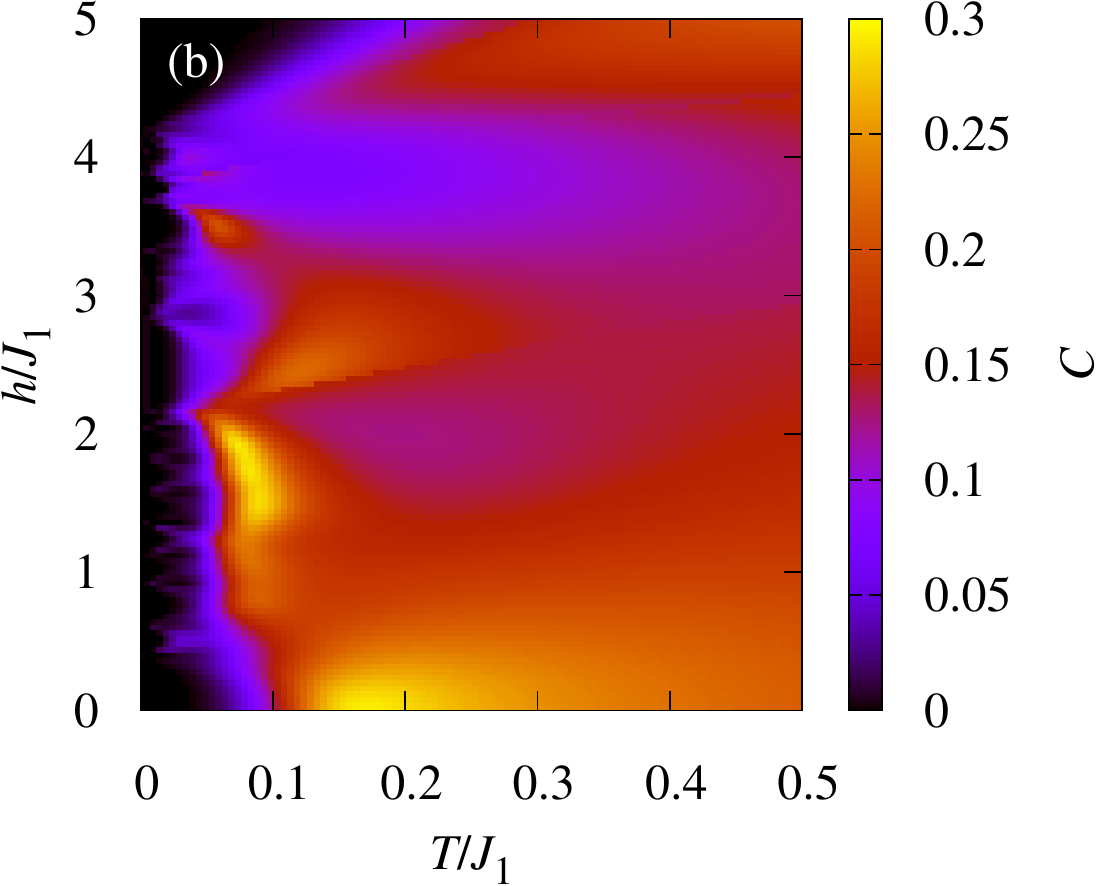} \\
\includegraphics[width=0.95\columnwidth]{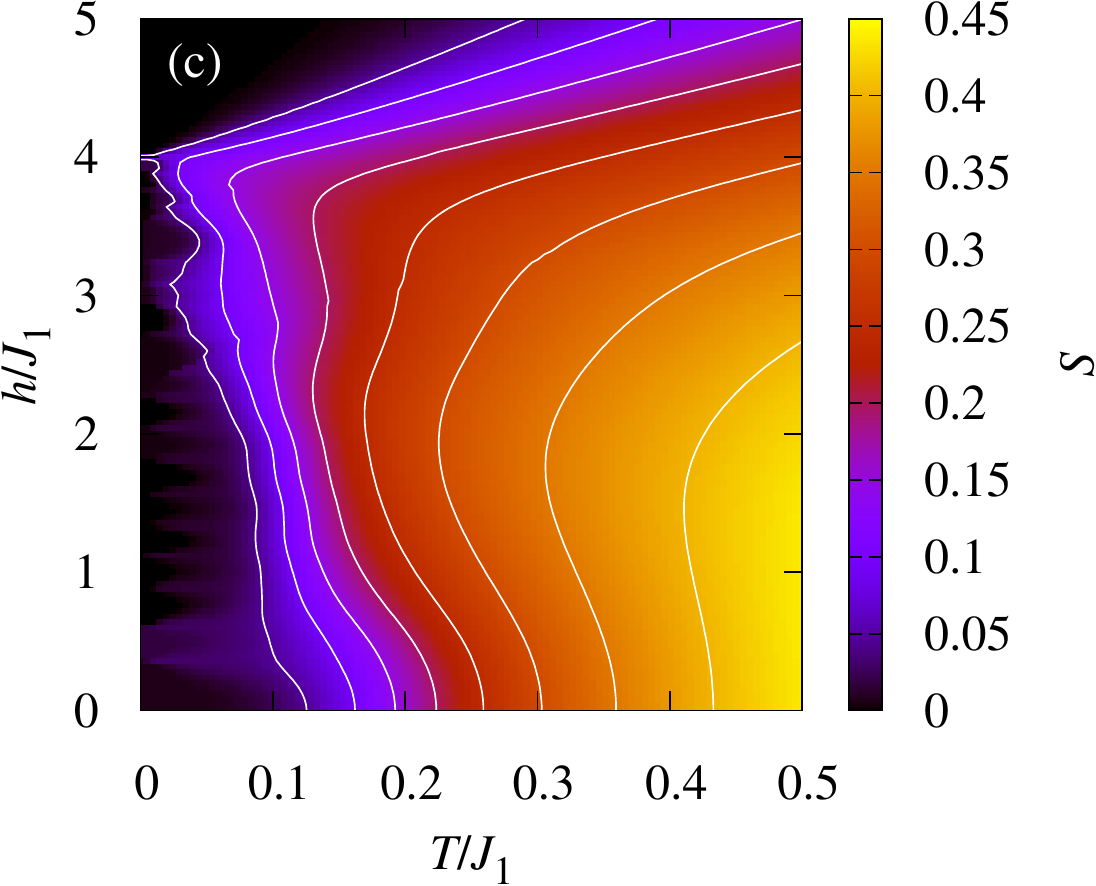}\qquad%
\includegraphics[width=0.95\columnwidth]{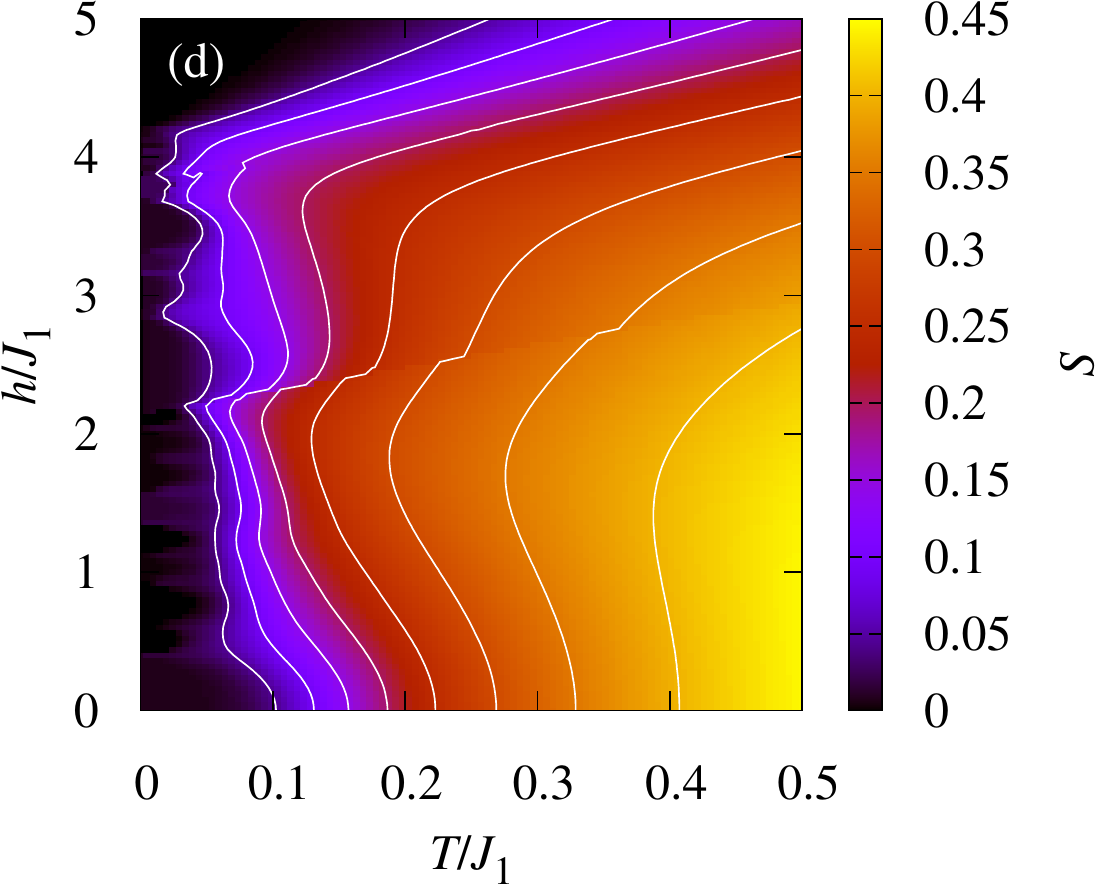}
\end{center}
\caption{Specific heat per spin $C$ (a,b) and entropy per spin $S$ (c,d)
as a function of temperature and magnetic field for $J_2/J_1=0.5$ (a,c)
and $0.55$ (b,d).
These figures are composites of different system sizes, starting with
$N=40$ for $m \lesssim 1/2$, going over $N=48$ for $1/2 \lesssim m \lesssim 1$,
$N=64$ for $m \gtrsim 1$ at high $T$ to $N=400$ for $m \gtrsim 1$ at low $T$.
The white solid lines in panels (c,d) trace values of constant entropy
per spin. The leftmost curve is for $S=0.05$, and $S$ increases in steps of $0.05$
with increasing temperature.
}
\label{fig:qThermo}
\end{figure*}

Finally, in Fig.~\ref{fig:qThermo} we show temperature-field scans
of the specific heat per spin $C$ (upper row of panels) and the entropy per spin $S$ (lower row of panels)
for $J_2/J_1=0.5$ (left column of panels)
and $J_2/J_1 = 0.55$ (right column of panels).
Here we have exploited the fact that the Hilbert space dimension decreases with increasing $m$
to build a composite of different system sizes such that we start with $N=40$ at low fields and
increase $N$ with increasing $m$ and decreasing temperatures such that we reach $N=400$
for $m\gtrsim1$ at low temperatures.

According to Subec.~\ref{sec:m1o2}, there should be no $m=1/2$ plateau at $T=0$ for $J_2/J_1=0.5$.
If the case $s=1/2$ is analogous to the classical case shown in Fig.~\ref{Diagram},
the maximum of $C$ at $T/J_1 \approx 0.1$ for $1 \lesssim h/J_1 \lesssim 2$
observed in Fig.~\ref{fig:qThermo}(a) might nevertheless indicate a transition out of
an $uuud$ phase stabilized by thermal fluctuations.
For $J_2/J_1=0.55$, we find an $m=1/2$ plateau in the range of fields
$2.24 \lesssim h/J_1 \lesssim 2.60$, see Fig.~\ref{fig:M}(c).
The maximum of $C$ for $T/J_1 \lesssim 0.1$ that one observes
in this range of fields in Fig.~\ref{fig:qThermo}(b) probably corresponds to a
transition out of this plateau state, as we have discussed in the context of Fig.~\ref{fig:Ch}(b).
Invoking again the analogy with the classical case Fig.~\ref{Diagram},
the continuation of this maximum to lower fields $1.2 \lesssim h/J_1 \lesssim 2.2$
and at $T/J_1 \lesssim 0.08$ might again correspond to a transition out of an $uuud$
state stabilized by thermal fluctuations. This suggests an intriguing scenario
where the crossover from stabilization by quantum to thermal fluctuations around
$h/J_1 \approx 0.55$ would be accompanied by a minimum in the transition temperature.

In the low-field limit, the specific heat of Fig.~\ref{fig:qThermo}(a,b) exhibits
a maximum around $T/J_1 \approx 0.2$. This could be just a fluctuation maximum.
However, it could also be a signature of a finite-temperature phase
transition for a plaquette or columnar valence-bond solid ground state as
suggested, e.g., in Refs.~\cite{doretto2014,Sheng2017,Qian2024,Huang2024,qiao2025}.
Such a state would break lattice symmetries (translations and/or rotations)
and could thus support a finite-temperature phase transition. However,
the features observed at $h=0$ in Fig.~\ref{fig:qThermo}(a,b) are broad.
If they correspond to phase transitions, finite-size effects would thus
be large. In any case, clarification of the exact meaning of these low-field low-temperature
maxima in the specific heat is a
question that goes beyond the scope of the present work.

\subsubsection{Entropy and magnetocaloric effect}

We now turn to the entropy $S$ that is shown in Fig.~\ref{fig:qThermo}(c,d).
This quantity and the related magnetocaloric effect were investigated previously
in the classical case \cite{Seabra2009} and for spin 1/2
\cite{Schmidt06,Schmidt07a,Schmidt07b,Schmidt2017}. However, these previous investigations
of the spin-1/2 model were restricted to $N\le 24$ and consequently to relatively high
temperatures.
Since the entropy is related to the specific heat $C/T$ by an integral, the information content
encoded in the entropy is equivalent to the specific heat that we discussed previously.
In this respect, the $S=0.05$ isentrope, {\it i.e.}, the leftmost white curve in Fig.~\ref{fig:qThermo}(c,d) can be considered to roughly trace the expected transition into the $uuud$
phase that we have discussed above. However, the entropy also yields a more direct
access to the magnetocaloric effect, as the isentropes show the behavior of the system
under adiabatic changes of the magnetic field. We observe distinct field-controlled
cooling around the saturation field in Fig.~\ref{fig:qThermo}(c,d) and close to
the quantum-phase transitions \cite{Zhu2003,Wolf2014} terminating the $m=1/2$ plateau in Fig.~\ref{fig:qThermo}(d).

Geometric frustration is expected to enhance the magnetocaloric effect \cite{Mike2003,Zhitomirsky2004,Honecker2006,Wolf2014}.
In the present case, we expect such an enhancement specifically at $J_2/J_1=1/2$
where the single-magnon spectrum becomes degenerate, recall Eq.~(\ref{omegaKhalf}).
Such an enhancement was indeed observed in previous works
\cite{Schmidt06,Schmidt07a,Schmidt2017} that scanned the ratio $J_2/J_1$
at a fixed temperature.
Here, such an enhancement is observed as a pronounced minimum in the lowest three
isentropes $T(h)$ around $h_{\rm sat}$ in Fig.~\ref{fig:qThermo}(c).

\section{Conclusion}
\label{sec:Concl}

We have revisited the behavior of the $J_1$-$J_2$ FSAFM in a magnetic field, pushing
previous investigations \cite{Honecker2000a,Honecker2000b,Honecker2002} to larger system sizes
and higher accuracy.

We first determined the phase diagram of the classical Heisenberg model at $J_2/J_1=1/2$.
Here, we confirmed order-by-disorder stabilization of the $uuud$ state around half of the
saturation field and for temperatures $0 < T \lesssim J_1/10$. In addition, we showed
that 2:1:1 and 3:1 supersolid phases emerge below and above the $uuud$ one, respectively.
Furthermore, we provided evidence that the $uuud$ phase persists in a small window
around $J_2/J_1=1/2$, although the full phase diagram of the classical FSAFM remains
to be investigated for $J_2/J_1\ne1/2$.

Then we turned to the spin-1/2 case. First, we presented zero-temperature magnetization
curves for several ratios of $J_2/J_1$ and provided evidence for the emergence
of a distinct $m=1/2$ plateau that can be understood as an $uuud$ state stabilized by quantum
fluctuations. We find a relatively large  stabilization regime $0.52 \lesssim J_2/J_1 \lesssim 0.7$
that is shifted to values $J_2 > J_1/2$
such that analytic treatment of this $uuud$ $m=1/2$ plateau requires expansion around
a classically unstable state \cite{Coletta2013}.
Intriguingly, the $uuud$ state is found to be stabilized by quantum fluctuations
in the regime $h \gtrsim h_{\rm sat}/2$, whereas the stabilization by thermal
fluctuations takes place at lower fields $h \lesssim h_{\rm sat}/2$.
Such a  trend is also known  for the Heisenberg antiferromagnet on the triangular lattice, see, e.g., discussion
in Ref.~\cite{Gvozdikova2011}.
We expect that the supersolid phases found in the classical case below and above the $uuud$ phase exist also for $s=1/2$.
While this expectation has been formulated earlier \cite{Honecker2002}, a detailed
investigation goes beyond the scope of the present work. Nevertheless, we mention
that Ref.~\cite{Shibata2016} found several non-trivial ordered states in the
magnetization curves of the $s=1/2$ FSAFM, although it is not entirely clear to us
if their Y-like and V-like phases match our 2:1:1 and 3:1 states below and above the
plateau, respectively.
Another point that might merit further attention is the question of the existence of the $m=1/3$ plateau predicted by a Chern-Simons theory \cite{Chern2002}. A systematic numerical investigation would require a sequence of system sizes divisible by three, but here we have only two of those at our disposal, namely $N=36$ and $48$.

Lastly, we computed finite-temperature properties of the $s=1/2$ FSAFM in a magnetic field,
relying mainly on FTLM.
We found preliminary signatures in the specific heat of a finite-temperature phase transition into the $uuud$
state. Like in the classical case, these finite-temperature ordering transitions remain restricted to low temperatures $T/J_1 \lesssim 0.1$.
We further observed that the degeneracy in the single-magnon spectrum at $J_2/J_1=1/2$ leads to an enhanced
magnetocaloric effect at the saturation field.

\begin{acknowledgments}
Johannes Richter was a prolific scientist. Regrettably, he did not witness completion of all of his projects. This is one of those that remained unfinished. Johannes Richter contributed the zero-temperature Lanczos calculations, part of the finite-temperature Lanczos results, and participated in the early stages of preparing the manuscript.

We are grateful to Jörg Schulenburg for providing {\tt spinpack} \cite{web:spinpack}, the code with which many of the computations were performed, and to Jürgen Schnack for sharing his insights on the finite-temperature Lanczos method.

The $N=36$ TPQ computations were performed with a precursor
of XDiag \cite{Wietek2025} using sublattice coding \cite{Wietek2018}.
\end{acknowledgments}

\bibliography{J1J2square}

\end{document}